\documentclass[journal]{IEEEtran}
\usepackage{color}
\usepackage{multirow}
\usepackage{mwe}
\usepackage{float}
\usepackage{xurl}
\usepackage{nccmath}
\usepackage{subcaption}
\usepackage{booktabs}
\usepackage{mwe}
\usepackage{cite}
\usepackage{graphicx}
\usepackage{textcomp}
\usepackage[normalem]{ulem}
\usepackage{amsthm,amsmath,amssymb,amsfonts}
\usepackage[linesnumbered,ruled]{algorithm2e}
\usepackage{algpseudocode}
\usepackage{xspace}
\usepackage{kotex}
\usepackage{soul}
\usepackage{mathtools, bbm}
\usepackage{xcolor,cancel}
\usepackage{tabulary}
\usepackage{array} 
\usepackage{mathtools}

\usepackage{hyperref}
\definecolor{linkcolour}{rgb}{0,0.2,0.6}
\definecolor{xgreen}{rgb}{0.2,0.6,0.0}
\definecolor{xred}{rgb}{0.7,0.1,0.0}
\hypersetup{colorlinks,breaklinks,citecolor=red!50, urlcolor=linkcolour, linkcolor=xgreen}
 
\usepackage{float}
\usepackage{physics}
\usepackage{bm}
\usepackage{amsthm, amsmath, amssymb}
\usepackage{thmtools}
\usepackage{thm-restate}
\usepackage{amssymb}
\usepackage{qcircuit}
\usepackage{tikz}
\usepackage{array}
\usepackage{xcolor}
\usepackage{mathtools}

\newcommand{\psib}{\ensuremath{\bm{\psi}}}

\newcommand{\BfPara}[1]{{\noindent\bf#1.}\xspace}

\def\BibTeX{{\rm B\kern-.05em{\sc i\kern-.025em b}\kern-.08em
    T\kern-.1667em\lower.7ex\hbox{E}\kern-.125emX}}

\begin{document}
\title{Quantum Multi-Agent Actor-Critic Neural Networks for Internet-Connected Multi-Robot Coordination in Smart Factory Management}
\author{Won Joon Yun,~\IEEEmembership{Graduate Student Member,~IEEE}, Jae Pyoung Kim, 
Soyi Jung,~\IEEEmembership{Member,~IEEE},\\ 
Jae-Hyun Kim,~\IEEEmembership{Member,~IEEE}, and
Joongheon Kim,~\IEEEmembership{Senior Member,~IEEE}
\thanks{A preliminary version of this paper was presented at the IEEE Int'l Conf. on Distributed Computing Systems (ICDCS), Bologna, Italy, July 2022~\cite{icdcs22yun}.}
\thanks{This research was funded by National Research Foundation of Korea (2022R1A2C2004869, 2021R1A4A1030775). \textit{(Corresponding authors: Soyi Jung, Jae-Hyun Kim, Joongheon Kim)}.}
\thanks{Won Joon Yun, Jae Pyoung Kim, and Joongheon Kim are with the School of Electrical Engineering, Korea University, Seoul 02841, Republic of Korea (e-mails: \{ywjoon95,paulkim436,joongheon\}@korea.ac.kr).}
\thanks{Soyi Jung and Jae-Hyun Kim are with the Department of Electrical and Computer Engineering, Ajou University, Suwon 16499, Republic of Korea (e-mails: \{sjung,jkim\}@ajou.ac.kr).}
}

\maketitle

\begin{abstract}
As one of the latest fields of interest in both academia and industry, quantum computing has garnered significant attention. Among various topics in quantum computing, variational quantum circuits (VQC) have been noticed for their ability to carry out quantum deep reinforcement learning (QRL). This paper verifies the potential of QRL, which will be further realized by implementing quantum multi-agent reinforcement learning (QMARL) from QRL, especially for Internet-connected autonomous multi-robot control and coordination in smart factory applications. However, the extension is not straightforward due to the non-stationarity of classical MARL. To cope with this, the centralized training and decentralized execution (CTDE) QMARL framework is proposed under the Internet connection. 
A smart factory environment with the Internet of Things (IoT)-based multiple agents is used to show the efficacy of the proposed algorithm. 
The simulation corroborates that the proposed QMARL-based autonomous multi-robot control and coordination performs better than the other frameworks.
\end{abstract}

\section{Introduction} 
In various Industry 4.0 scenarios, automated and autonomous management of smart factory systems are getting a lot of attention nowadays~\cite{iot103,8310596,iot101,iot102,9699416,9650739,8950429}. 
For the automation of factory management, the use of autonomous multiple mobile robots is widely studied~\cite{iot104,9247159,9216961}. According to the Verizon Report~\cite{verizon}, \textit{Industry 4.0 is squarely underway in manufacturing. The global market is expected to reach \$219.8 billion by 2026, and autonomous mobile robots are becoming key workhorses in this transformation}. To realize the efficient and effective autonomous multi-robot control and coordination, multi-agent reinforcement learning (MARL)-based algorithms are essentially required~\cite{hu2020voronoi,9682599}.

Recently, revolutionary innovations have been made in distributed learning and MARL due to the remarkable evolution in computing hardware and deep learning algorithms~\cite{9682599}. 
Moreover, the developments in quantum computing hardware and algorithms placed further emphasis on this trend~\cite{schuld2022quantum}, resulting in the incentivization of the research on quantum machine learning. 
Nowadays, quantum machine learning is at a newborn level compared to conventional machine learning.
For instance, in the classification task of quantum machine learning, the performance of quantum machine learning is low given the MNIST dataset at 32.5\% of top-1 accuracy \cite{hanruiwang2022quantumnas} on quantum computers, and 74.2\% \cite{yun2022pvm} on ideal quantum machines. 
However, the theoretically discovered advantages (\textit{i.e., quantum supremacy}) are being experimentally proven recently~\cite{Cornelissen22STOC,wiedemann2022quantum,arute2019quantum}.
The potential of quantum algorithms is evident from their ability to downsize the model parameters while maintaining accuracy by exploiting quantum entanglements~\cite{oh2020qcnn-simple}.
In addition, the empirical result of \cite{jerbi2021variational} shows that quantum machine learning outperforms the empirical result of classical machine learning.
An outstanding example of this is the variational quantum circuit (VQC) architecture, also known as a quantum neural network (QNN)~\cite{ijcnn21hong,icufn21kwak}. 
QNN is a quantum circuit that reproduces the function of a classical deep neural network. By combining the QNN and classical deep learning models, hybrid quantum-classical models are built, which allow QRL to be carried out. Compared to RL, QRL uses lesser model parameters but significantly reduces the training and inference time~\cite{chen2020variational,ictc21kwak} while consuming lesser computing resources as well~\cite{carleo2019machine}.
Thus, it is clear that quantum machine learning using quantum computing will become a big trend in the near future.
 This paper aims to combine VQC with the classical MARL to extend QRL to quantum MARL (QMARL).

The agents in the MARL environment interact with each other by either cooperating or competing. This interaction is realized based on Internet-of-Things (IoT)-based connectivity technologies. 
These interactions result in a non-stationary reward for each agent, which hinders the convergence of MARL training. The centralized training and decentralized execution (CTDE) method is used~\cite{arxiv2017_VDN} to deal with the non-stationarity of the MARL model. 
In this scenario, the reward is distributed to all agents concurrently by concatenating their state-actions pairs. A na\"ive implementation of a VQC version of CTDE is possible, as shown in~\cite{icdcs22yun}. However, such implementation causes the qubits to increase with the number of agents because when QRL is carried out via VQC, the state-action pairs are represented by qubits. Consequently, quantum errors will also increase with the qubits~\cite{shor1995scheme}, significantly affecting the MARL convergence and scalability.  Furthermore, quantum error correction is not yet viable in the current noisy intermediate-scale quantum (NISQ) era.

This paper intends to improve on various pre-existing methods of implementing VQCs, agent policies, and state encoding~\cite{yun2022quantum,chen2020variational,lockwood2020reinforcement}. Three significant differences exist for our proposed VQC compared to previous works, which are parameter sharing, non-random VQC design, and 2-variables dense encodings. Firstly, parameter sharing refers to sharing model parameter values between agents.  All the agents in previous works had individual, distinct policies meaning more agents required more policies, resulting in excess computing power consumption in the process of formulating them. In this improved model, there will only be one policy that will be shared among the agents, increasing computing power efficiency. The second improvement is the non-random VQC design. The VQCs  used in previous works are composed of randomly selected quantum gates. Although the performance is remarkable, it cannot be easily reproduced because of its randomness. The same model might not show the same performance in another iteration because of the random quantum gates. However, this is improved in this paper by designing a fixed model and removing the random nature of the previous VQC. This ensures the reproducibility and stability of the model. Finally, the proposed model in this paper utilizes the 2-variables dense encoding method instead of the 4-variables dense encoding method. The original encoding method is capable of reducing the dimensions of given data. Although this may be good for the NISQ-era quantum circuits, it inevitably causes a loss of information. The proposed 2-variables dense encoding method does not reduce data dimensions, but it is still compatible with NISQ-era quantum circuits. Thus, information loss is prevented, which will improve the performance of this model.

\BfPara{Contributions} 
The major contributions of this research are  summarized as follows.
\begin{itemize}
\item This paper first provides a quantum-based MARL solution for autonomous multi-robot control and coordination in smart factory applications.
\item An improved and novel CTDE QMARL framework which utilizes parameter sharing on policy, VQC design, and 2-variables dense encodings is additionally proposed.
\item Lastly, via extensive experiments, the proposed QMARL framework is proven to be superior to the classical MARL model by carrying out simulations in smart factory scenarios. The results show that the proposed model produces higher performance than the others.
\end{itemize}

\BfPara{Organization} 
The rest of this paper is organized as follows. The preliminaries of this paper are described in Sec.~\ref{sec:2}. Our considering autonomous mobile robots coordination for smart manufacturing is described in Sec.~\ref{sec:3}. 
Sec.~\ref{sec:4} introduces our proposed algorithm; and the numerical results and demonstration of the proposed algorithm are shown in Sec.~\ref{sec:5}.
Sec.~\ref{sec:6} concludes this paper and presents future work.
Note that the notations in this paper are listed in Table \ref{tab:notation}. Most equations and notations used here are based on the \textit{Dirac} notations used in~\cite{simeone2022introduction}.

\begin{table}[t!]
\normalsize%\small
    \caption{List of notations}
    \label{tab:notation}
    \centering
    \begin{tabular}{c|l}
        \toprule[1pt]
        \multicolumn{2}{c}{\textsf{Scenario Notations}}\\\midrule
        $N$ & The number of AMR agents\\
        $M$ & The number of sites/warehouses\\
        $T$ & An episode length\\
        $o^n$ & The observation of $n$-th AMR agent\\
        $a^n$ & The action of $n$-th AMR agent\\
        $\textbf{a}$ & The actions set of AMR agents, i.e.,  $\textbf{a}=\{a_n\}^N_{n=1}$. \\
        $s$ & The ground truth state\\
        $q^c_{m,t}$ & The load status of $m$-th warehouse at time $t$\\
        $q^e_{n,t}$ & The load status of $n$-th AMR agent at time $t$\\
        $q^e_{\max}$ & A load capacity of warehouse \\
        $q^e_{\max}$ & A load capacity of AMR agent  \\
        \midrule
        \multicolumn{2}{c}{\textsf{Quantum Computing Notations}}\\\midrule
        $|\psi\rangle$ & Entangled quantum state\\
        $\langle O \rangle$ & Observable\\
        $\Gamma$ & Pauli-$\Gamma$ gate, e.g., $\Gamma \in \{X,Y,Z\}$\\
        $R_\Gamma $ & Rotating $\Gamma$ gate, e.g., $\Gamma \in \{X,Y,Z\}$\\
        $(\cdot)^\dagger$ & Complex conjugate operator\\
        % $f_{\textit{Enc}}$ & State encoding circuit\\
        % $f_{\textit{VQC}}$ & Variational quantum circuit\\
        $\mathcal{M}$ & Measurement operator\\
        \bottomrule[1pt]
     \end{tabular}
\end{table}

\section{Preliminaries of Quantum Computing}\label{sec:2}

\BfPara{Single Qubit Quantum State} 
QC utilizes a \textit{qubit} as the basic unit of computation. The qubit represents a quantum superposition state between two basis states, denoted as $|0\rangle$ and $|1\rangle$. There are two ways to describe a qubit state,
\begin{equation}
\label{eq:qubit}    
|\psi\rangle = \alpha|0\rangle + \beta|1\rangle,
\end{equation}
where $\|\alpha\|_2^2 + \|\beta\|_2^2 = 1$, as well as,
\begin{equation}
\label{eq:bloch}    
|\psi\rangle = \cos\left(\frac{\delta}{2}\right)|0\rangle + e^{i\varphi}\sin\left(\frac{\delta}{2}\right)|1\rangle,
\end{equation}
where $\delta \in [-\pi,\pi]$ and $\varphi \in [-\pi,\pi]$. The former is based on a normalized 2D complex vector, while the latter is based on polar coordinates $(\delta,\varphi)$ from a geometric viewpoint.
The qubit state is mapped into the surface of a 3D unit sphere (\textit{Bloch sphere}).
In addition, a quantum gate is a unitary operator transforming a qubit state into another qubit state, which is represented as a $2\times2$ matrix with complex entries. 
The single-qubit Pauli gates $X$, $Y$, and $Z$ are defined as follows,
\begin{equation}
X = \begin{bmatrix}
0 & 1 \\ 1 & 0
\end{bmatrix},\quad Y = \begin{bmatrix} 0 & -i \\ i & 0 \end{bmatrix},\quad  Z = \begin{bmatrix} 1 & 0 \\ 0 & -1 \end{bmatrix}.
\end{equation}

There are additional quantum gates that are frequently used, i.e., $R_{\text{x}}$, $R_{\text{y}}$, and $R_{\text{z}}$.
These are rotation operator gates rotate a single qubit by $\delta$ around their corresponding axes in the Bloch sphere and the single qubit operation can be expressed as the following equations,
\begin{equation}
 R_X(\delta) = e^{-i \frac{\delta}{2} X}, \quad R_Y(\delta) = e^{-i \frac{\delta}{2} Y}, \quad R_Z(\delta) = e^{-i \frac{\delta}{2} Z},
 \end{equation}
where rotation angles are denoted as $\delta\in\mathbb{R}[0,2\pi]$.
These basic Pauil and rotation gates are unitary matrices, $U^\dag U = I$, where $I$ denotes an identity matrix. 

\BfPara{Multi-Qubit Quantum State} 
Multi-qubit system enables super-fast quantum computing due to quantum superposition. The well-known quantum algorithms (e.g., Shor algorithm \cite{Shor94} and Grover search \cite{Grover96STOC}) are based on the multi-qubit system. The quantum state with $q$ qubits is denoted as $|\psib\rangle = |\psi_1\rangle \otimes |\psi_2\rangle \otimes \cdots \otimes |\psi_q\rangle = \sum^{2^q-1}_{n=0} \alpha_n |n \rangle$, where $\otimes$, $\alpha_n$ and $|n\rangle$ stand for superposition operator (i.e., tensor-product), and $n$-th probability amplitude and $n$-th basis of $q$-qubits quantum state, respectively. Note that the sum of squared magnitude of probability amplitude equals 1, \textit{i.e.}, $\sum^{2^q-1}_{n=0} |\alpha_n|^2 = 1$ \cite{simeone2022introduction}.
To realize quantum superposition, there are quantum gates that operate on multiple qubits, called controlled rotation gates. They act on a qubit according to the signal of several control qubits, which generates quantum entanglement between those qubits. Among them, the \textit{Controlled}-$X$ (or CNOT) gate $\textit{CX} =\begin{bmatrix}I & 0 \\ 0 & X\end{bmatrix}$ is one of the widely used control gates. These multi-qubit gates allow quantum algorithms to work with their features on VQC, which will eventually be utilized for QMARL.

\section{Autonomous Mobile Robots Coordination for Smart Manufacturing}\label{sec:3}
\begin{figure*}[t!]
\!\!\!\!\!\!\!\!\!\!\!\!\!\!\!\includegraphics[width=1.1\textwidth]{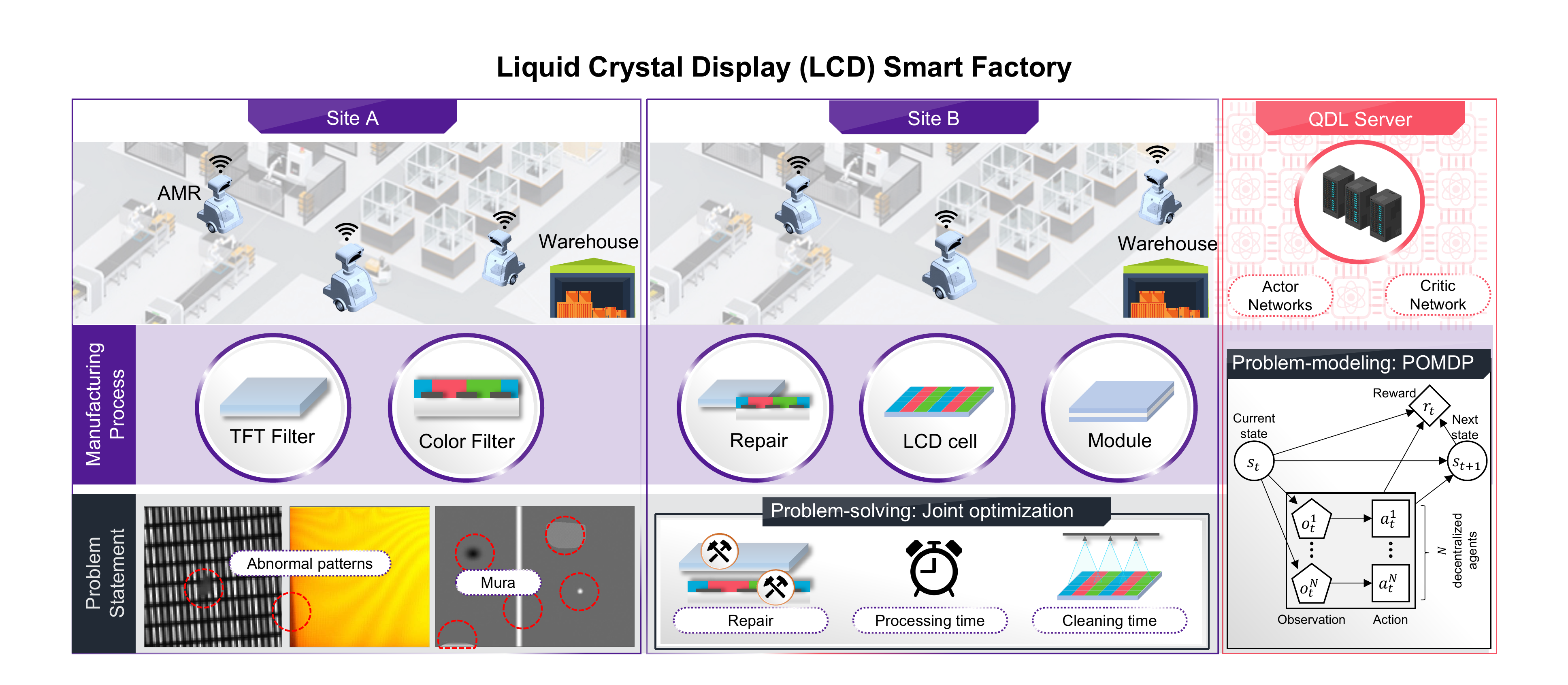}
    \caption{System model: Liquid crystal display panel manufacturing using quantum multi-agent reinforcement learning}
    \label{fig:1}
\end{figure*}
% \begin{figure}[t!]
% \centering
% \includegraphics[width=.8\columnwidth]{IOTJ_Figures/iotj-pomdp.pdf}
% \caption{Decentralized POMDP for LCD smart factory modeling.}
% \end{figure}
\subsection{Design of an Autonomous Mobile Robot System}
An automated guided vehicle (AGV) is a portable robot that travels along lines or wires marked on the floor or navigates using radio waves, vision cameras, magnets, or lasers. AGVs are widely used in industrial applications to transport heavy materials around large industrial facilities such as factories and warehouses. Therefore, it is obvious that AGVs are essential to smart factory management. 
The autonomous mobile robot (AMR) differs from AGV because it has various sensors that enable autonomous location identification and search by detecting surrounding static and dynamic objects. Their paths are generated based on static and dynamic obstacles in real time so that AMR can travel freely without a predefined path. While the system is more flexible, real-time path generation poses additional challenges that fleet management systems (FMSs) must deal with such as, performing activities, e.g., shipping transport orders, routing vehicles, and scheduling task execution. Note that AMRs are tightly combined, which leads to high computational complexity. 
For example, the performance of AMRs suffer when considering all possible AMR paths, even though the numbers of AMRs and transfer orders are relatively small. As a result, centralized AMR fleet management and order execution optimization are often not performed in real time. Therefore, the use of MARL algorithms is widely considered and studied~\cite{8988635}.
For further performance improvement, QMARL can be additionally utilized, as we discuss in this paper.

\subsection{Automated LCD Smart Factory with Multiple AMRs}
Thanks to the properties of QC, QC has shown that QC could save many orders of magnitude in energy consumption compared to classical supercomputers~\cite{villalonga2020establishing}. Regarding QRL, recent studies show quantum supremacy~\cite{jerbi2021variational}.
In this paper, we consider a liquid crystal display (LCD) smart factory system which utilizes DC-based AMRs. 
As shown in Fig.~\ref{fig:1}, the color thin-film transistor (TFT) LCD panel consists of two glass substrates; a TFT array substrate and a color filter substrate. TFT LCD panels are fabricated by a combination of five processes; TFT array filter process, color filter process, repair process, cell fabrication process, and module assembly process. The first two processes (i.e., TFT array filter and color filter processes) are carried out at Site A, the two substrates are carried by AMR and the rest of the process is carried out at 
Site B. Each AMR has a role in the transition from providing services to flexible areas that require decisions to be made based on dividing the service area into several zones. 
In the process of manufacturing TFT LCDs, various defective LCDs can occur which should not be used. Therefore, to prevent the usage of such defective LCDs, the AMRs must identify the defective products and request a quality verification of the LCD. Techniques of detecting defects among LCDs have already been developed and implemented in smart factories~\cite{LCD-REFs,lee2017robust,LCDTFT}. By using the \textit{precision} parameter proposed by the works above, the AMRs will recognize defective LCDs and unload them in another collection point dedicated for defects.

In this paper, we assume that all AMRs have the optimal trajectory planners and charging schedulers such as \cite{xue2017trajectory,tvt202106jung}. Thus, our proposed QMARL model must plot the trajectory of each AMR such that the defective LCDs are separated while the normal products are properly unloaded.
For communication, the QDL server is wire-linked to every site, and each site is wirelessly connected with AMRs. Since the packet size is small, and the transmit power is sufficient in LCD smart factory, we assume that the packet loss is negligible. 
The QDL server receives observation from AMRs, reconfigures the state, and finally transmits action decisions to AMRs.
For flexible manufacturing, AMRs should be properly planned to load goods, unload goods, and control the quality of LCD. Moreover, the decision-making process of scheduling and dispatching these resources is essential for optimal utilization and high AMR productivity performance.

\BfPara{Problem Definition and Formulation} 
In this situation, a quantum deep learning (QDL) server supports the decision-making process for efficiently scheduling material handling systems, under the fundamental concept of the CTDE-based QMARL framework. Specifically, the QDL server makes distributed and sequential decisions for each AMR to determine their goods (i.e., the number of goods to carry in each AMR and requesting quality control)  for eliminating the overflow and underflow of delivering goods in each AMR.

\section{Quantum Multi-Agent Actor-Critic Network for Autonomous Multi-Robot Coordination}\label{sec:4}

\subsection{Fundamental MDP Formulation}\label{sec:4-1}
Our considering autonomous multi-robot coordination in a smart factory environment consists of $M$ sites and $N$ AMR agents. The smart factory environment is mathematically modeled with POMDP (referred to as Sec.~\ref{sec:2}).
Hereafter, we explain the description based on $m$-th site, $n$-th agent, and time step $t$.

\subsubsection{Load Dynamics} 
Each site has a warehouse where the load capability is $c^{\text{W}}_{\max}$. In addition, the load capacities of AMR agents is under the maximum capacities $c^{\text{A}}_{\max}$. 
AMR agents receive goods (e.g., LCD panels or TFTs) from other AMRs. In this paper, we denote the load weights $b^{\text{A},n}_t$.
The load weights follow the uniform distribution $\forall b^{\text{A},n}_{t} \sim \mathcal{U}(0, w_{\textit{load}} \cdot b_{\max})$.
The warehouse and AMR agents have loading status $c^{\text{W},m}_t$ and $c^{\text{A},n}_t$ that are temporally loaded goods. All AMR agents carry their goods to warehouses. 
The dynamics are as follows,
\begin{equation} 
    c^{\xi,n}_{t+1} = \mathit{clip}(c^{\xi,n}_{t}- a^{\xi,n}_{t} 
    + b^{\xi,n}_{t},0,c^{\xi}_{\max}), 
\end{equation}
where $\xi\in\{\text{W},\text{A}\}$ identifies the warehouse and an AMR agent. The terms $a^{\xi,k}_{t}$ and $b^{\xi,n}_{t}$ imply the total delivered goods weights and the received goods weights of $m$-th warehouse or $n$-th AMR agent, respectively. Note that $a^{A,n}_{t}$ is $n$-th AMR agent's action. In addition, a clipping function is defined as $\mathit{clip}(x,x_{\min},x_{\max}) \triangleq \min(x_{\max}, \max(x,x_{\min}))$. 

\subsubsection{Quality Control}
We assume that the loads have been classified by the previous defect detection process. 
In the defect detection process, four types are given to the loads (i.e., true positives, false positives, false negatives, and true negatives),  which make the statistics (\textit{i.e.}, precision, recall, and F-score). 
This paper considers the load status as the quality statistics (\textit{e.g.}, precision). 
Among them, quality statistics are given to AMR agents. The AMR agents can make action decisions for re-requesting quality control of the load. If the loads are requested for quality control, the loads undergo quality verification by quality engineers. We assume that quality engineers can detect all defects on load perfectly. However, the quality re-assurance process via quality engineers additionally requires $\tau_{\textit{qual}}$ time delay. 

\subsubsection{Utility Design}
We design the utility for quality, time delay, and load balancing. First of all, AMR agents receive the goods with the type of true positives ($\textit{TP}$) and false positives ($\textit{FP}$). The metric, i.e., precision, can represent the ratio of positive predictive value, which is written as follows:
\begin{equation}
    u^{q,n}_t = \frac{\textit{TP}^n_t}{\textit{TP}^n_t + \textit{FP}^n_t}, 
\end{equation}
where $\textit{TP}^n_t = \sum_{\textit{load} \in l^n_t} \mathbbm{1}(\textit{load} = \textit{TP})$ and $\textit{FP}^n_t  = \sum_{\textit{load} \in l^n_t} \mathbbm{1}(\textit{load} = \textit{FP})$ stand for the true positives and false positives of $n$-th AMR, respectively. Note that $l^n_t$ denotes the whole load defect status of $n$-th AMR. 
Regarding the delay, we measure the processing time. Thus, the delay utility of $n$-th AMR agent is written as follows: 
\begin{equation}
    u^{d,n}_t = - \Big(1 + \tau_{\textit{qual}} \cdot q^n_t \Big),
\end{equation}
where $q^n_t$ denotes the quality control action. 
If $q^n_t=1$, the loads are conveyed to quality engineers; otherwise, the loads are conveyed to other-site. 
Finally, load balancing is to minimize the total amount of overflowed load and the event where the load is empty. Thus, the utility for load balancing is written as follows:
\begin{align}
u^{b,n}_t &= \mathbbm{1}_{(c^{A,n}_{t+1} = 0)}\cdot \tilde{c}^{A,n}_{t} + \mathbbm{1}_{(c^{A,n}_{t+1} = c^A_{\max})}\cdot \hat{c}^{W,n}_{t}\\
u^{W,m}_t &= \mathbbm{1}_{(c^{W,m}_{t+1} = 0)}\cdot \tilde{c}^{W,m}_{t} + \mathbbm{1}_{(c^{W,m}_{t+1} = c^W_{\max})}\cdot \hat{c}^{W,m}_{t}
    \label{eq:reward}
\end{align}
 where $\tilde{c}^{W,m}_{t} = |c^{W,n}_{t}-a^{W,n}_{t}+b^{W,n}_{t}|$ and $\hat{c}^{W,n}_{t} = |c_{\max} - \tilde{c}^{W,n}_{t}|$.
Note that $r(s_t,\mathbf{a}_t) \in [-\infty, 0]$ (negative) because this paper considers the occurrence of abnormal loading status (e.g., load overflow or underflow) as a negative utility.
The objective is to maximize the total precision and minimize the total delay and overflowed or underflowed event.

\subsection{POMDP Setup} \label{sec:4-2}
This subsection introduces the formal definition of POMDP, i.e., a stochastic decision-making model under uncertainty among agents~\cite{Springer2016_POMDP}; and our proposed QMARL is mathematically modeled with this fundamental concept of POMDP. Note that POMDP is defined as a tuple $\langle \mathcal{N}, \mathcal{S}, \mathcal{A}, P, r, \mathcal{Z}, O, \rho, \gamma, T\rangle$. The sets of states and observations are represented as $\mathcal{S}$ and $\mathcal{Z}$, respectively. $\mathcal{N}:= \{1, \cdots, N\}$ and $s\in\mathcal{S}$ denote the set of $N$ agents and the current state of the environment, respectively. The initial state $s_0 \sim \rho$ follows the distribution $\rho$. The action of $n$-th agent $a^n \in \mathcal{A}$ is discrete or continuous actions, and the joint action is denoted as $\mathbf{a}:= \{a^n\}_{n=1}^N$. The transition is determined with probability function $P(s'|s, \mathbf{a}):\mathcal{S}\times\mathcal{A}\times\mathcal{S}\rightarrow\mathcal{S}$, where $s'$ denotes the next state. The shared reward $r_t = r(s_t, \mathbf{a}_t):\mathcal{S}\times\mathcal{A}\rightarrow \mathbb{R}$ is given to whole agents.
In Dec-POMDP, the true state $s$ is not directly given to agents. Each agent $n \in \mathcal{N}$ has observation $z^n \in \mathcal{Z}$ from observation function $O(s,a): \mathcal{S} \times \mathcal{A} \rightarrow \mathcal{Z}$.
We consider that all agents have parameter-shared policy denoted as $\pi_\theta$. 
Thus, the policy $\pi_\theta$ takes the $n$-th agent's observation $z^n_t\in \mathcal{Z}$ and decides $n$-th agent's action as $\pi_\theta(a|z^n): \mathcal{Z} \times \mathcal{A} \rightarrow [0,1]$. The objective of POMDP is to obtain the optimal policy $\pi_\theta^* = \arg\max_{\pi_\theta} \mathbb{E}_{\mathbf{a}\sim\pi_\theta}[\sum^T_{t=1}\gamma^{t-1} \cdot r_t$], where $\gamma \in \mathbb{R}[0,1]$ and $T$ denote the discount factor and finite time, respectively. Based on this definition, we design the POMDP as follows:

\subsubsection{Observation} Each AMR agent partially obtains its observation. Because the parameter shared policy $\pi_\theta$ is used, the observation contains the binary indicator vector $\textbf{n}_2$. In addition, $n$-th AMR agent makes its action decision with its loading status $c^{A,n}_t$, and the current loading status of warehouse  $\{c^{W,m}_t\}_{m=1}^M$. In summary, $n$-th agent's observation is defined as $z^n_t \triangleq \{\textbf{n}_2, c^{A,n}_{t}\}\cup\{c^{W,m}_{t}\}^M_{m=1}$.
\subsubsection{State} A state information variable containing information about all AMR agents' and warehouse's loading status is designed. The state variable at time $t$ is as $s_t = \{c^{A,n}_t, u^{d,n}_t\}^{N}_{n=1} \cup 
\{c^{W,m}_t\}^M_{m=1} $. Note that the state information is utilized as the input of the quantum critic network.
\subsubsection{Action} It is considered that AMR agents can choose which warehouse to convey goods, where the destination space is defined as $\mathcal{I} \triangleq \{1, \cdots, M\}$.  
In addition, AMR agents can determine conveying quantity to the warehouse. 
The conveying quantity and the quality control space is defined as $\mathcal{P}\triangleq \{p_{\min}, \cdots, p_{\max}\}$ and $\mathcal{Q}\triangleq \{0, 1 \}$, respectively. 
Finally,  $n$-th AMR agent's action and its action space are defined as $a^n_t:= (i^n_t, p^n_t, q^n_t) \in \mathcal{A} \equiv \mathcal{I}\times \mathcal{P} \times \mathcal{Q}$.
\subsubsection{Reward}
The objective of POMDP is to minimize the total amount of overflowed load and the event where the load is empty. Thus, the reward $r(s_t, \mathbf{a_t})$ is defined as follows, 
\begin{equation}
r(s_t, \mathbf{a}_t) = \sum^{N}_{n=1}(u^{q,t}_t + w_d \cdot u^{d,n}_t + w_b\cdot u^{b,n}_t) + w_W\cdot \sum^M_{m=1}u^{W,m}_t,
\end{equation}
where $w_d$, $w_b$, and $w_W$ stand for reward coefficients for time delay, and load-balancing of AMR agents and sites, respectively.

\subsection{Quantum Multi-Agent Actor-Critic Network Design}\label{sec:4-3}
\begin{figure}[t!]
    \centering
    \includegraphics[width=1\columnwidth]{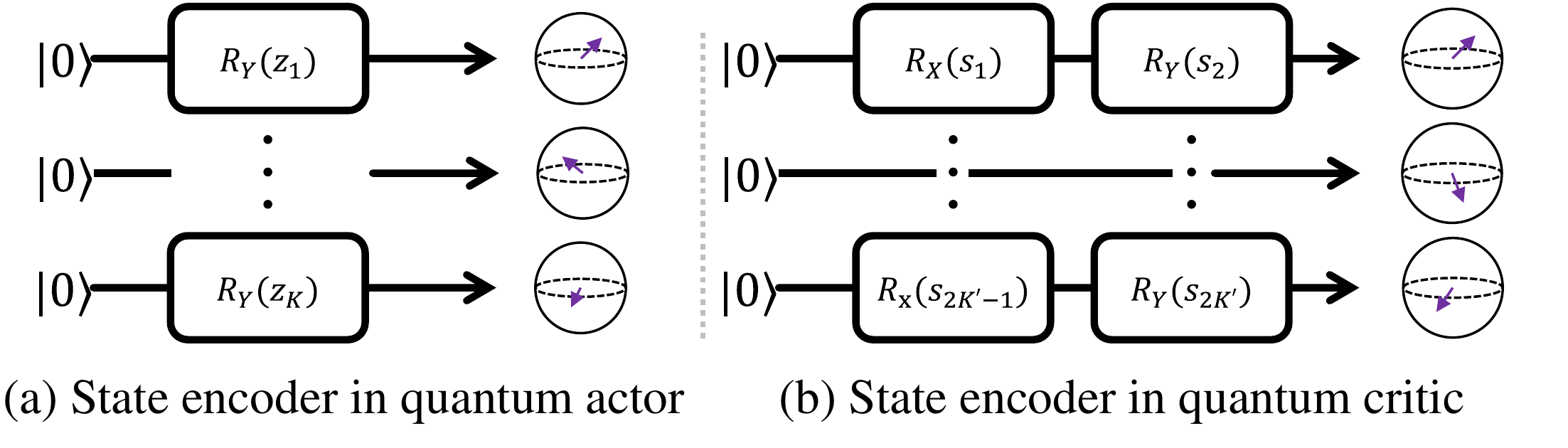}
    \caption{The illustration of the state encoder.}
    \label{fig:2}
\end{figure}

\subsubsection{State Encoding Circuit}
The state encoding circuit is leveraged for feedforwarding a state input. Fig.~\ref{fig:2} presents the two schemes of state encoder. Fig.~\ref{fig:2}(a)/(b) need a single gate or two gates per qubit, respectively. 
Despite the encoding system showing the best performance when the number of qubits is equal to the number of input variables, the number of input variables in RL (i.e., state) must be larger than the number of qubits \cite{ictc21kwak}.
Thus, this paper considers two state encoders under the consideration of the environment, as follows,
\begin{align}
    U^a_{\textit{\textit{enc}}}(z) &= \Big[\otimes^{K}_{k=1}(R_{Y}(x^z_{k}))\Big] \cdot |0\rangle^{\otimes q_{\textit{actor}}},\\
    U^c_{\textit{enc}}(s) &= \Big[\otimes^{K'}_{k'=1} (R_{Y}(x^s_{2k'}) \cdot R_{X}(x^s_{2k'-1}))\Big] \cdot |0\rangle^{\otimes q_{\textit{critic}}},
\end{align}
where $x^z_k$ and $x^s_{k'}$ stand for $k$-th entry of observation $z$ and $k'$-th entry of state $s$, respectively.
Note that $U^a_{\textit{enc}}(z)$ and $U^c_{\textit{enc}}(s)$ denote an actor observation encoder and critic state encoder. The actor observation encoder and critic state encoder work in the $K$ and $K'$ qubits system, respectively.
\subsubsection{Parameterized Circuit and Quantum Measurement}
\begin{figure}[t!]
    \centering\includegraphics[width=1\columnwidth]{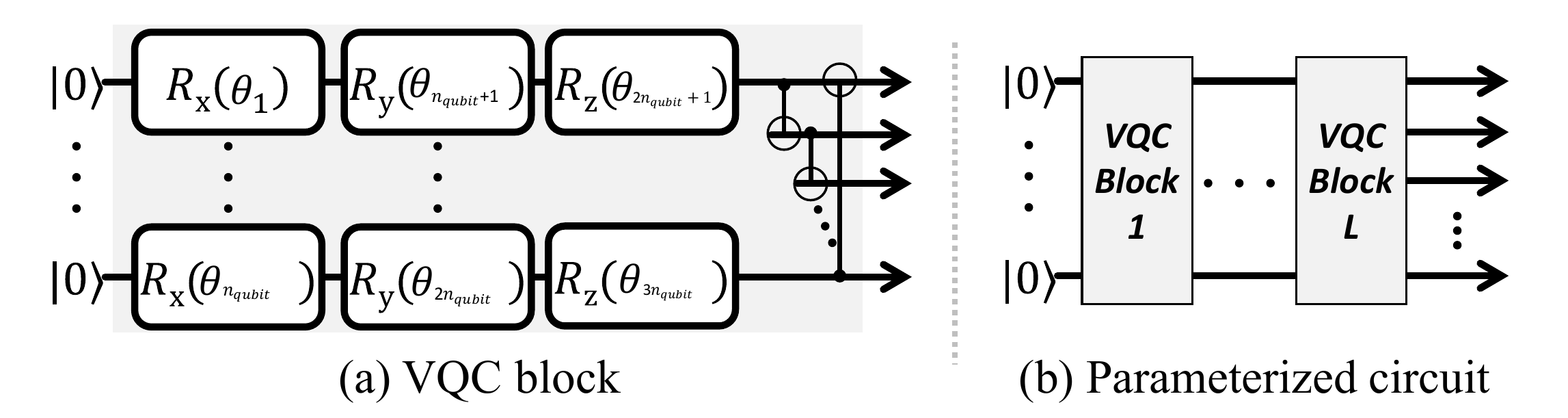}
    \caption{The illustration of the parameterized circuit.}
    \label{fig:3}
\end{figure}

A parameterized circuit is a quantum circuit that performs numerical tasks such as estimation, optimization, approximation, and classification using learnable parameters. As shown in Fig.~\ref{fig:3}(a), The VQC block consists of rotating gates with different directions and \textit{Controlled-Z} gate, i.e., $R_X$, $R_Y$, $R_Z$, and $\textit{CZ}=\begin{bmatrix} I & 0 \\ 0 & Z\end{bmatrix}$. Note that $\textit{CZ}$ is used to entangle qubits.
To improve the circuit's performance, this paper configures the parameterized circuit with multi-VQC blocks, which requires additional trainable parameters $\theta$ as shown in Fig.~\ref{fig:3}. To obtain the desirable outputs, the measurement $\mathcal{M}$ is leveraged, which calculates the expected value of superpositioned quantum states based on its computational basis.   
In summary, the observable (i.e., expected value) is written as follows:
\begin{align}
    \langle O\rangle_{x, \theta} &=\prod\limits_{M\in\mathcal{M}}\langle0|U^{\dagger}_{\textit{enc}}(x)U^{\dagger}_{\textit{VQC}}(\theta)MU_{\textit{VQC}}(\theta)U_{\textit{enc}}(x)|0\rangle,
\end{align}
where $\langle O\rangle_{x, \theta}$ is the output of VQC with inputs $x$ and circuit parameter $\theta$; $\mathcal{M}$ is the set of quantum measurement bases in VQC with $|\mathcal{M}| \leq n_{qubit}$.

\subsubsection{Implementation on Quantum Actor-Critic}  
The proposed QMARL for a smart factory in this paper is decentralized for scalability. Every AMR agent in the QMARL has a VQC-based policy, i.e., agents do not require communication among agents. The observables of the actor/critic are as follows,
\begin{align}
\!\!\!\!\!\langle O_a\rangle_{o, \theta} \!\!&=\!\left\{\!\langle0|U^{a\dagger}_{\textit{enc}}(o)U^{a\dagger}_{\textit{VQC}}(\theta)MU^a_{\textit{VQC}}(\theta)U^a_{\textit{enc}}(o)|0\rangle\!\right\}\!\!_{M\in\mathcal{M}_a},\label{obs:actor}\\
\!\!\!\!\!\langle O_c\rangle_{s, \phi} \!\!&=\!\left\{\!\langle0|U^{c\dagger}_{\textit{enc}}(s)U^{c\dagger}_{\textit{VQC}}(\phi)MU^c_{\textit{VQC}}(\phi)U^c_{\textit{enc}}(s)|0\rangle\!\right\}\!\!_{M\in\mathcal{M}_c}.\label{obs:critic}
\end{align}

\BfPara{Quantum Actor}\label{sec:qa}
For the quantum actor, the observable of \eqref{obs:actor} is used to calculate the probabilities of actions of each AMR agent. 
Then, the quantum policy is written via a softmax function of its observable,
\begin{align}
    \pi_{\theta}(a_t|{z}_t)&=\textit{softmax}(\beta_a \langle O_a\rangle_{{z}^n_t, \theta}),
\end{align}
where 
\begin{equation}
\textit{softmax}(\mathbf{x}) \triangleq \left[\frac{e^{x_1}}{\sum_{i=1}^N e^{x_i}};\cdots;\frac{e^{x_N}}{\sum_{i=1}^N e^{x_i}}\right]
\end{equation}
and $\beta_a$ is the scaling factor for an actor observable, respectively.
At the time $t$, the actor policy of $n$-th agent makes an action-decision with the given observation $o^n_t$, which is denoted as $\pi_{\theta}(a^n_t|o^n_t)$. Note that $\theta$ denotes parameters of $n$-th actor.
Then, the action $a^n_t$ is computed as follows,
\begin{equation}
    {a^n_t=\arg\max\limits_{a}\pi_{\theta}(a|o^n_t),}
\end{equation}
and note all agents use the same policy by parameter sharing.

\BfPara{Quantum Centralized Critic}\label{sec:qc}
The centralized critic is adopted for CTDE as a state-value function. At time $t$, the parameterized critic estimates the discounted returns given $a_t$ as follows:
\begin{multline}
    V_\phi(s_t) \!=\! \beta_c \langle O_c \rangle_{s_t, \phi} \!\simeq \!\mathbb{E}[\sum_{t'=t}^{T} \gamma^{t'-t}\!\!\cdot r(s_{t'},\!\mathbf{u}_{t'}) | s_t \!=\! s],
\end{multline}%  is 
where $\gamma$, $T$, $\mathbf{a}_t$, $\beta_c$, and $r(s_{t'},\mathbf{a}_{t'})$ stand for a discounted factor $\gamma \in [0,1)$, an episode length, the actions of all agents, scaling factor for a critic observable and reward functions that the state $s_{t'}$ and action $\mathbf{a}_t'$ are given, respectively. In addition, $\phi$ presents the trainable parameters of a critic. Here, $s_t$ is the ground truth state at $t$.

\begin{algorithm}[t!]
%\small
	\caption{Quantum Multi-AMR Agents Training}
	\label{alg:1}
		Initialize the critic and actor networks with weights $\theta$ and $\phi$ and the replay buffer $\mathcal{D} = \left\{ \right\}$\;
		Initialize the target networks as: $\phi^{\mathsf{T}} \leftarrow \phi$\;
		
    \For{episode = 1, MaxEpisode}{
        $\triangleright$ Initialize \textbf{Smart Factory Environments}\;
		$t = 0$\; $s_0=\text{initial state}$\;
    		\While{$s_t \neq terminal$ and $t < \text{episode limit}$}{
    		\For{each agent $n$}{
    		Calculate $\pi_{\theta}(a|o^n_t)$ and sample $a^n_t$\;
    		}
    		Get reward $r_t$ and next state and observations $s_{t+1}$, $\mathbf{o}_{t+1}= \{o^n_t\}^N_{n=1}$\;
    		$\mathcal{D} = \mathcal{D} \cup \left\{(s_t, \mathbf{o}_t, \mathbf{a}_t, r_t, s_{t+1}, \mathbf{o}_{t+1} )\right\}$\;
    		$t=t+1$\; 
    		}
		\For{each timestep $t$ in each episode in batch $\mathcal{D}$}{
		{Get $V_\phi(s_t)$; $V_{\phi^{\mathsf{T}}}(s_{t+1})$}\;
		{Calculate the target $y_t${ with \eqref{eq:td}}}\;
		}
		{Calculate $\nabla_\theta \mathcal{L}_{\textit{actor}}$, $\nabla_\phi\mathcal{L}_{\textit{critic}}$, 
		and update $\theta$, $\phi$\;}
		\If{$\text{target update period}$}{
		Update the target network, $\phi^{\mathsf{T}} \leftarrow \phi$}}
\end{algorithm}

\subsection{Training Algorithm}\label{sec:4-4}
The objective of MARL agents is to maximize discounted returns. To derive the gradients for the maximization objective, we leverage the joint state-value function $V_\phi$.
To train $V_\phi$, this paper leverages a multi-agent policy gradient (MAPG), which is formulated as follows,
\begin{eqnarray}
\nabla_{\!\theta} \mathcal{L}_{\textit{actor}} &=& -\mathbb{E}_{\textbf{a} \sim \pi_\theta}\! \left[ \sum\limits^{T}_{t=1}\!\sum\limits^{N}_{n=1} y_t \nabla_{\!\theta} \!\log\pi_{\theta}(a^n_t|z^n_t) \! \right]\!, \label{eq:l_actor}\\
\nabla_\phi \mathcal{L}_{\textit{critic}} &=& \nabla_{\!\phi}\!\sum^{T}_{t=1}\left\|y_t\right\|^2, \label{eq:l_critic}
\end{eqnarray}
subject to
\begin{equation}
y_t =  r(s_t,\mathbf{a}_t) + \gamma V_{\phi^\mathsf{T}}(s_{t+1}) - V_{\phi}(s_t), \label{eq:td} 
\end{equation}
where $\phi^{\mathsf{T}}$ is the parameters of target critic network. Note that \eqref{eq:l_actor} and \eqref{eq:l_critic} are for following the parameter-shift rule~\cite{crook19}, written as follows:
\begin{eqnarray}
    \frac{\partial \mathcal{L}_{\textit{actor}}}{\partial \theta_i}  &=& \frac{\partial \mathcal{L}_{\textit{actor}}}{\partial \pi_\theta} \!\!\cdot\!\! \frac{\partial \pi_\theta}{\partial \langle O \rangle_{o,\theta}}\!\!\cdot\!\!\Big[\langle O \rangle_{o,\theta \!+\!\! \frac{\pi}{2} \textbf{e}_i } \!\!-\!\! \langle O \rangle_{o,\theta \!-\! \frac{\pi}{2} \textbf{e}_i }\Big] , \\
    \frac{\partial \mathcal{L}_{\textit{critic}}}{\partial \phi_j}  &=& \frac{\partial\mathcal{L}_{\textit{critic}}}{\partial V_\phi} \!\!\cdot\!\! \frac{\partial V_\phi}{\partial \langle O \rangle_{s,\phi}}  \!\!\cdot\!\!\Big[\langle O \rangle_{s,\phi \!+\!\! \frac{\pi}{2} \textbf{e}_j } \!\!-\!\! \langle O \rangle_{o,\phi \!-\! \frac{\pi}{2} \textbf{e}_j }\Big],
\end{eqnarray}
where $\textbf{e}_i$ and $\textbf{e}_j$ stand the $i$- and $j$-th standard bases of parameterized vectors $\theta$ and $\phi$, respectively. Note that the two left partial derivatives are derived by classical computing, and the last term is obtained by quantum computing.
The detailed training procedure is presented in \textbf{Algorithm~\ref{alg:1}}.

\section{Performance Evaluation}\label{sec:5}

\begin{table}[t!]
\normalsize
%\small
\caption{The benchmark schemes.}
\centering
\begin{tabular}{lll}\toprule[1pt]
    \textbf{Schemes} & \textbf{Computing method} &\textbf{\# of parameters} \\
    \midrule
    \textsf{Proposed} & Quantum & $\approx110$\\
    \textsf{Comp1} &  Quantum/Classical & $\approx110$\\
    \textsf{Comp2} &  Classical & $\approx110$\\
    \textsf{Comp3} &  Classical & $\approx40$K\\
    \textsf{Comp4} &  Random Walk & None\\
    \bottomrule[1pt]
\end{tabular}
\label{tab:3}
\end{table}

\begin{table}[t!]
\normalsize
%\small
\caption{The experiment parameters.}
\centering
\begin{tabular}{@{}l|r@{}}\toprule[1pt]
    \textbf{Parameters} & \textbf{Values} \\
    \midrule
    The number of sites ($M$) & $2$ \\
    The number of AMRs ($N$) & $6$ \\
    The load capacity of warehouse & $2,000$\,kg\\
    The load capacity of AMR agent & $500$\,kg\\
    Observation dimension & $6$ \\
    Precision (Reported \cite{LCDTFT})      & $\{61.9, 95.8, 97.1\} \%$ \\
    Weight of TFT-LCD (Reported \cite{LCDweight}) & $6$\,kg\\
    Action dimension & $5$ \\
    State dimension & $8$ \\
    Episode length & $30$\,timestep\\
    Reward coefficient $(w_d,w_b,w_W)$ & $(0.1, 1, 10)$\\
   Time delay by quality engineers ($\tau_{\textit{qual}}$) & 3\,timestep\\\midrule
    %Hyper-parameters ($w_{\mathcal{P}}$, $w_\mathcal{R}$) & $(0.3, 4)$\\
    %Transmitted packets from the cloud ($b^{c,k}_t$) & $0.3$\\
    %The capacity of queue ($q_{\max}$) & $1$ \\
    Actor observable hyperparameter\,$\beta_a$ & $3$  \\
    Critic observable hyperparameter\,$\beta_c$ & $35$  \\
    Optimizer & Adam optimizer\\
    The number of gates in $U^a_{p}$ and $U^c_{p}$  & $54$ \\
    The number of qubits of actor  & $8$ \\
    The number of qubits of critic & $8$ \\
    Learning rate of actor & $1\times 10^{-2}$ \\
    Learning rate of critic & $1\times 10^{-3}$  \\
    Weight decay & $1\times 10^{-5}$  \\
    \bottomrule[1pt]
\end{tabular}
\label{tab:2}
\end{table}

\begin{figure*}[t!]
\centering
\includegraphics[width=.9\textwidth]{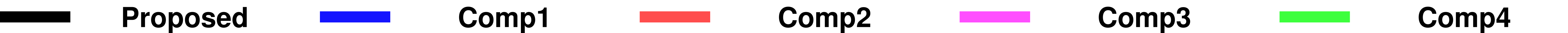}\\
\begin{tabular}{@{}p{.2\linewidth}@{}p{.2\linewidth}@{}p{.2\linewidth}@{}p{.2\linewidth}@{}p{.2\linewidth}@{}}
     \includegraphics[width=\linewidth]{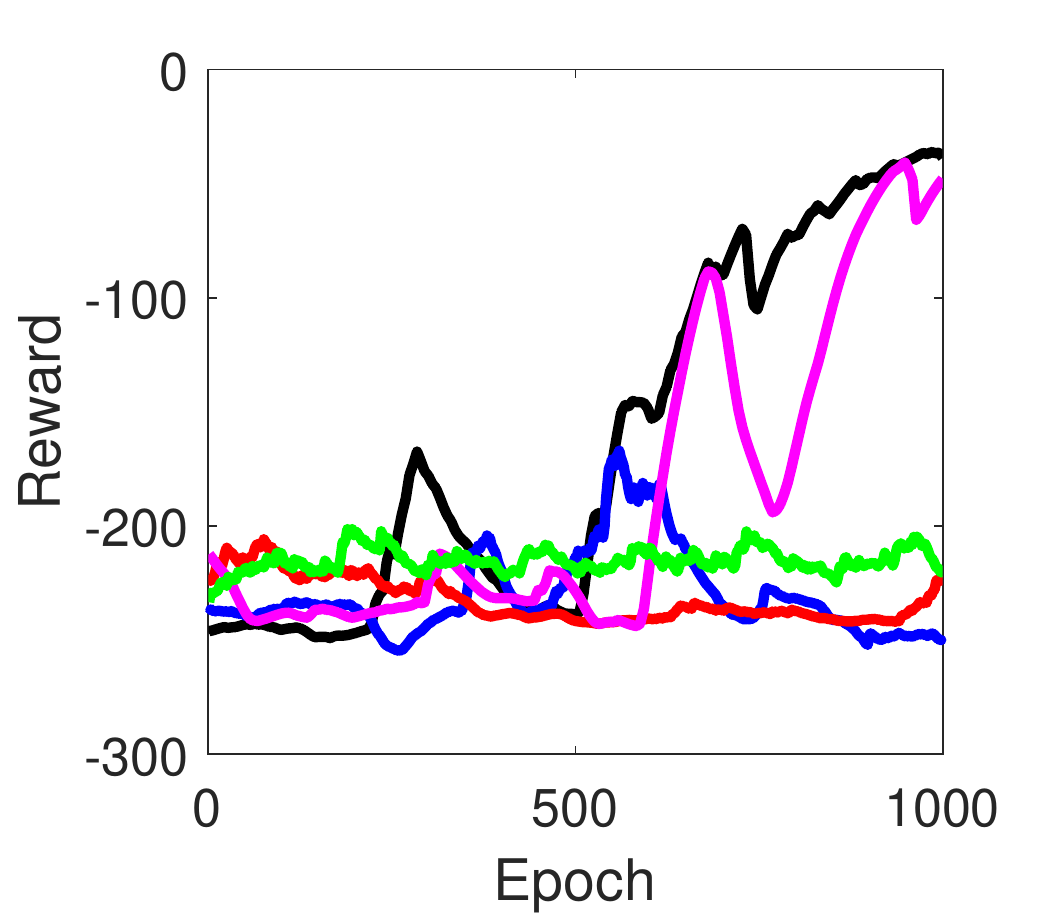} &
     \includegraphics[width=\linewidth]{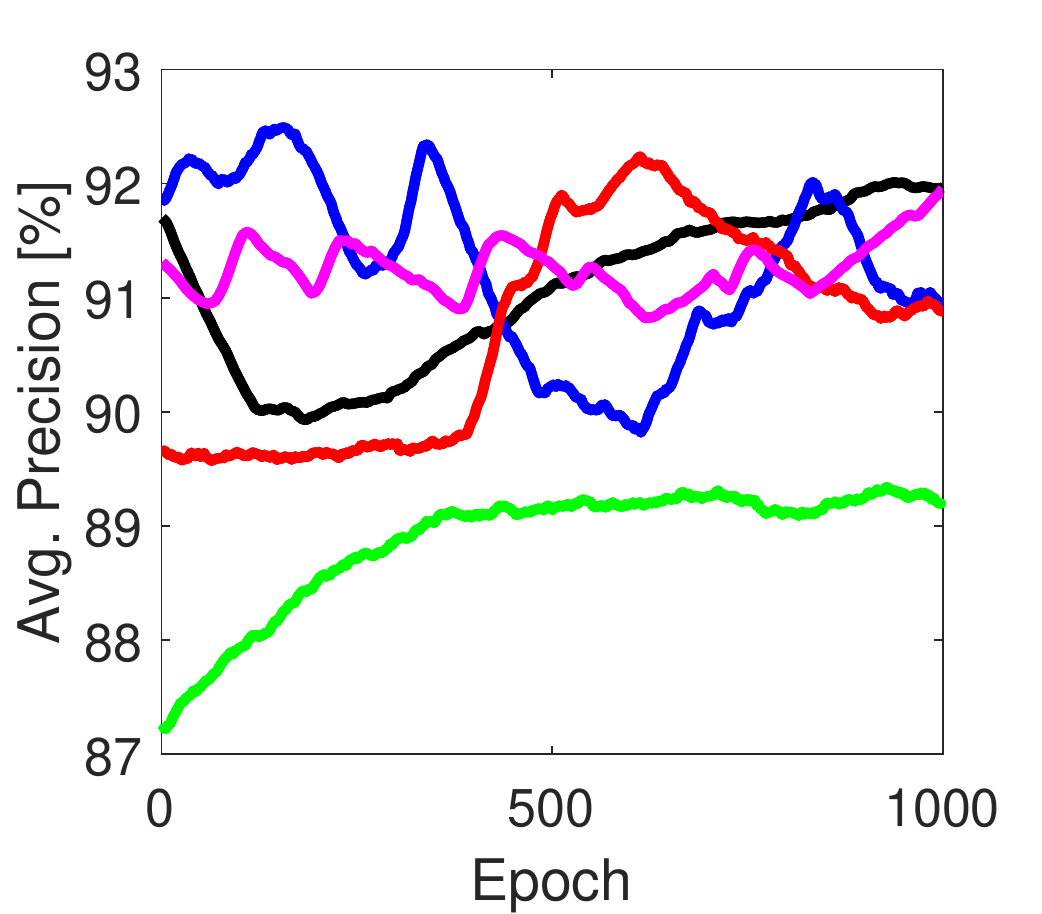} &
     \includegraphics[width=\linewidth]{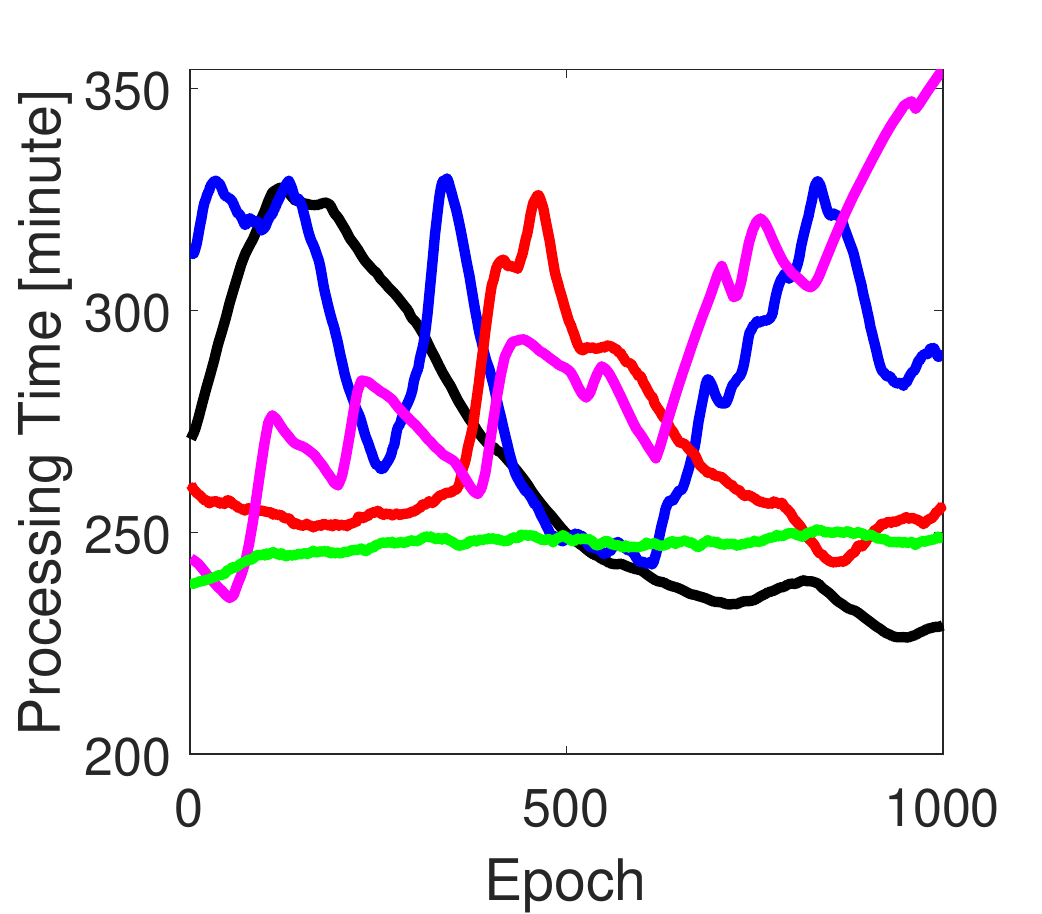} &
     \includegraphics[width=\linewidth]{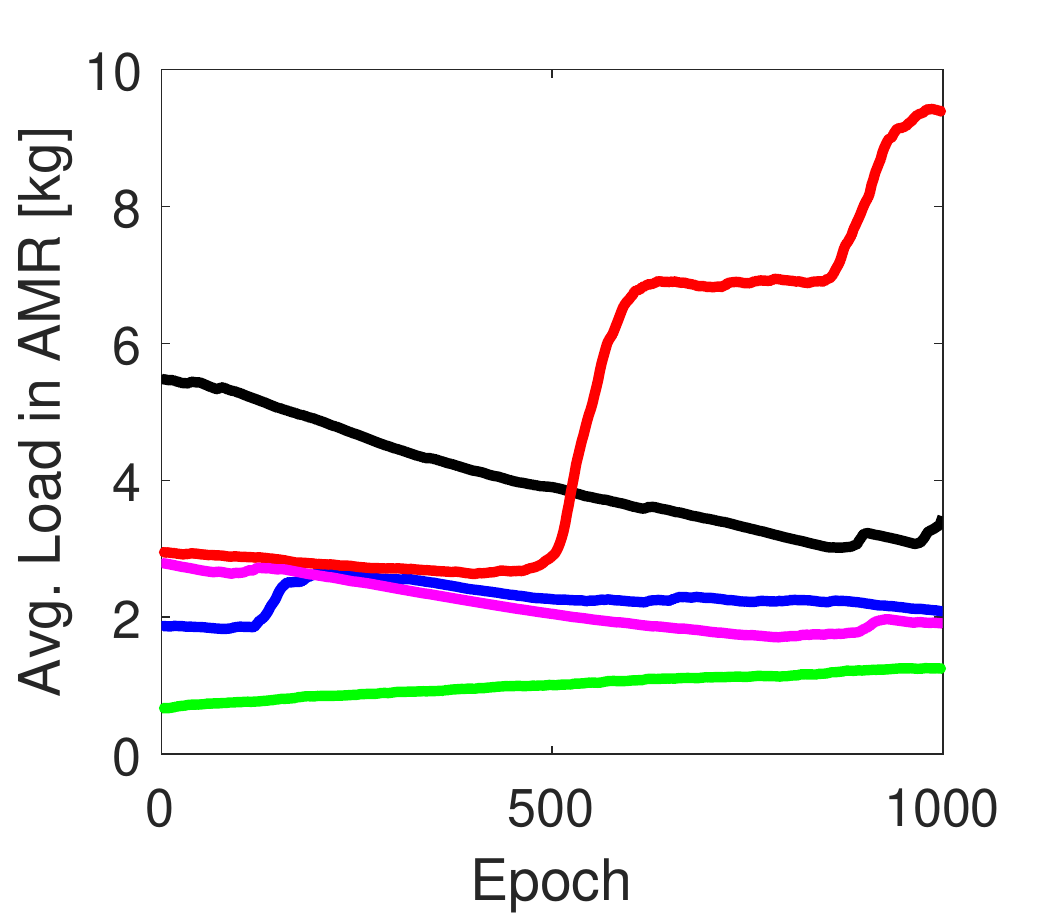}  &
     \includegraphics[width=\linewidth]{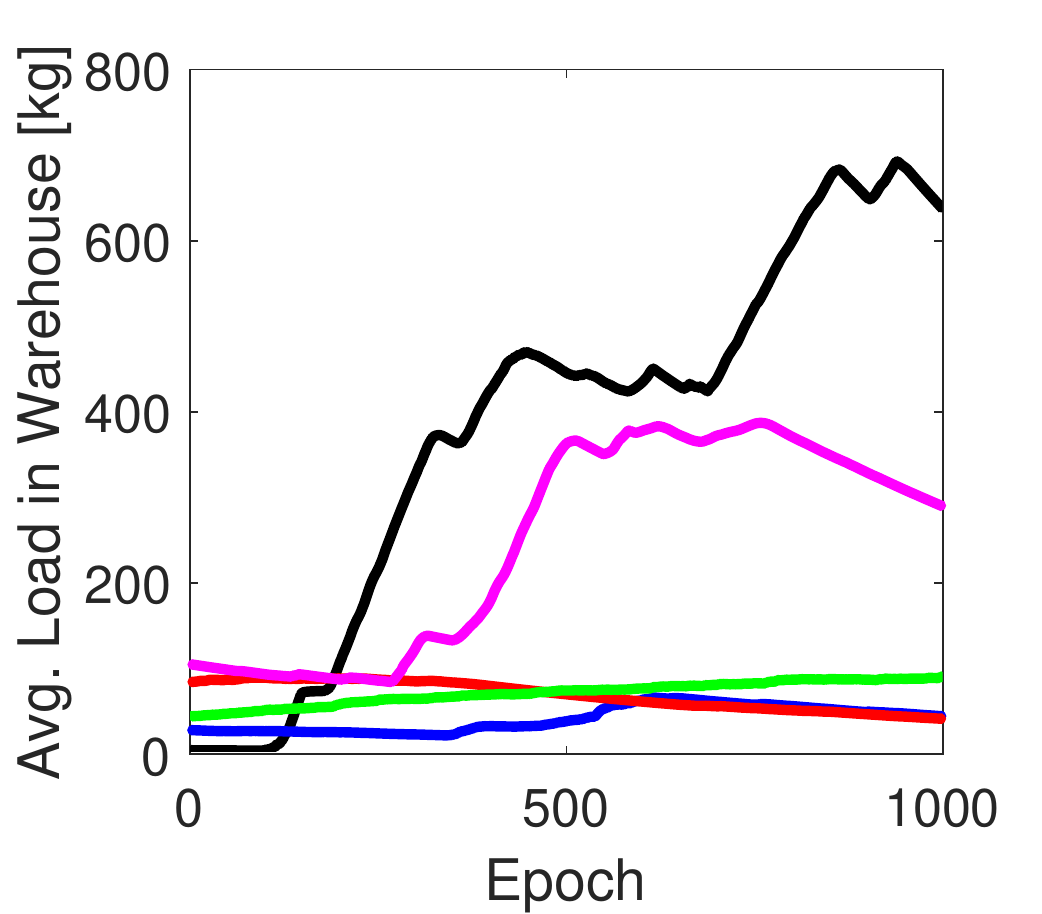} 
     \\
     \multicolumn{1}{c}{\scriptsize (a) Total reward} & 
     \multicolumn{1}{c}{\scriptsize(b) Precision}   &
     \multicolumn{1}{c}{\scriptsize(c) Total Processing Time}   &      
     \multicolumn{1}{c}{\scriptsize(d) Avg. loaded amount in AMR}  & 
     \multicolumn{1}{c}{\scriptsize(e) Avg. loaded amount in warehouse}  \\
     \end{tabular}
     \begin{tabular}{@{}p{.25\linewidth}@{}p{.25\linewidth}@{}p{.25\linewidth}@{}p{.25\linewidth}@{}}
     \includegraphics[width=\linewidth]{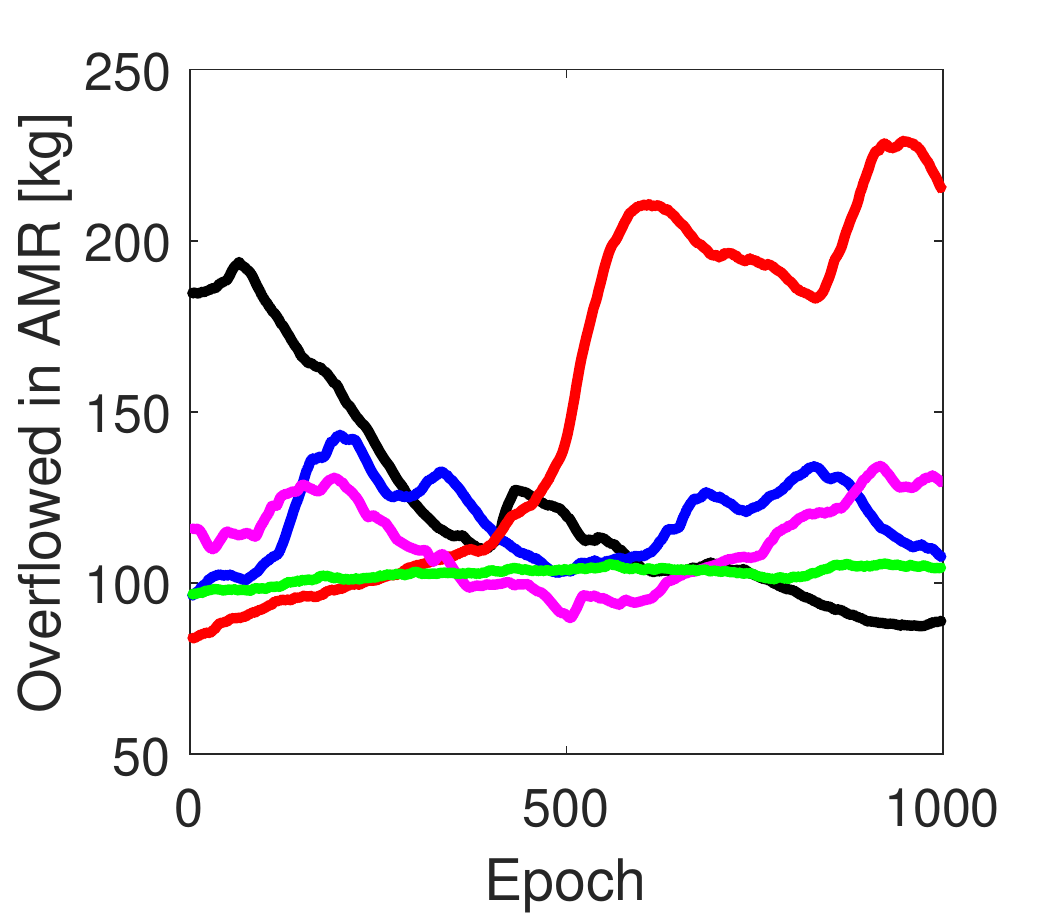} &
     \includegraphics[width=\linewidth]{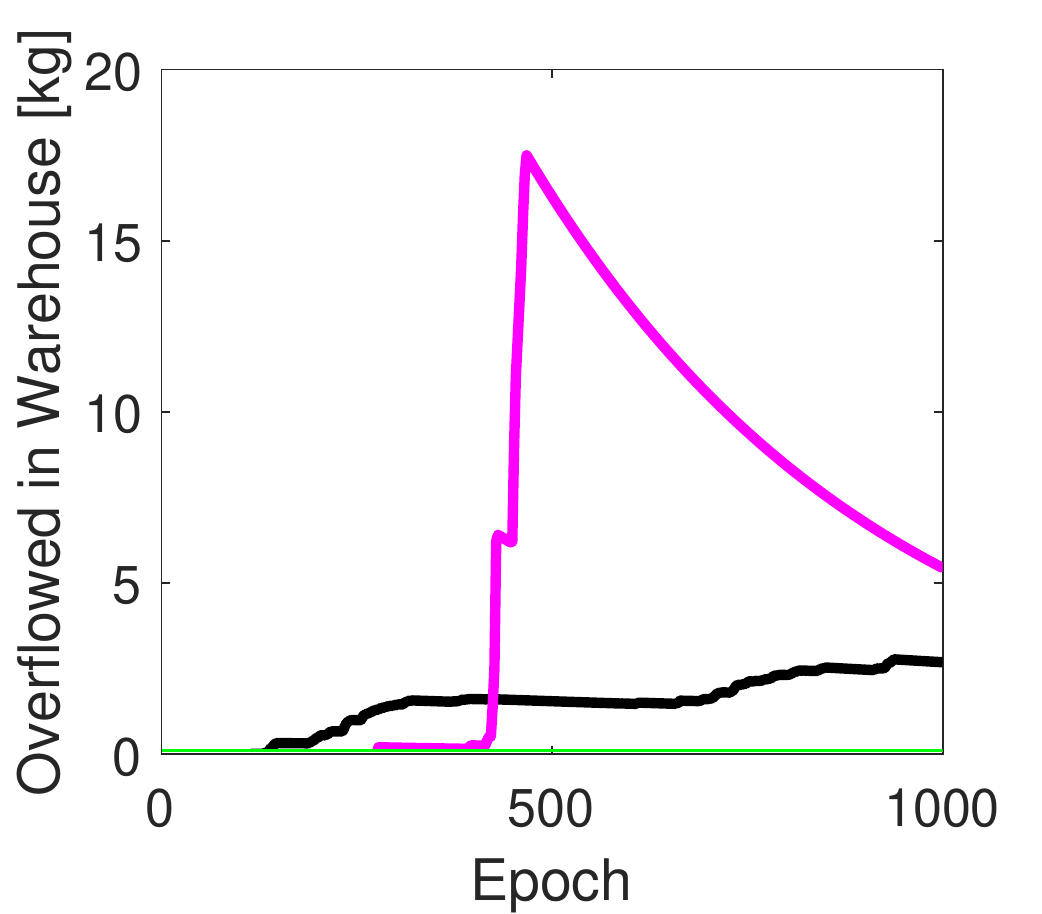}  &
     \includegraphics[width=\linewidth]{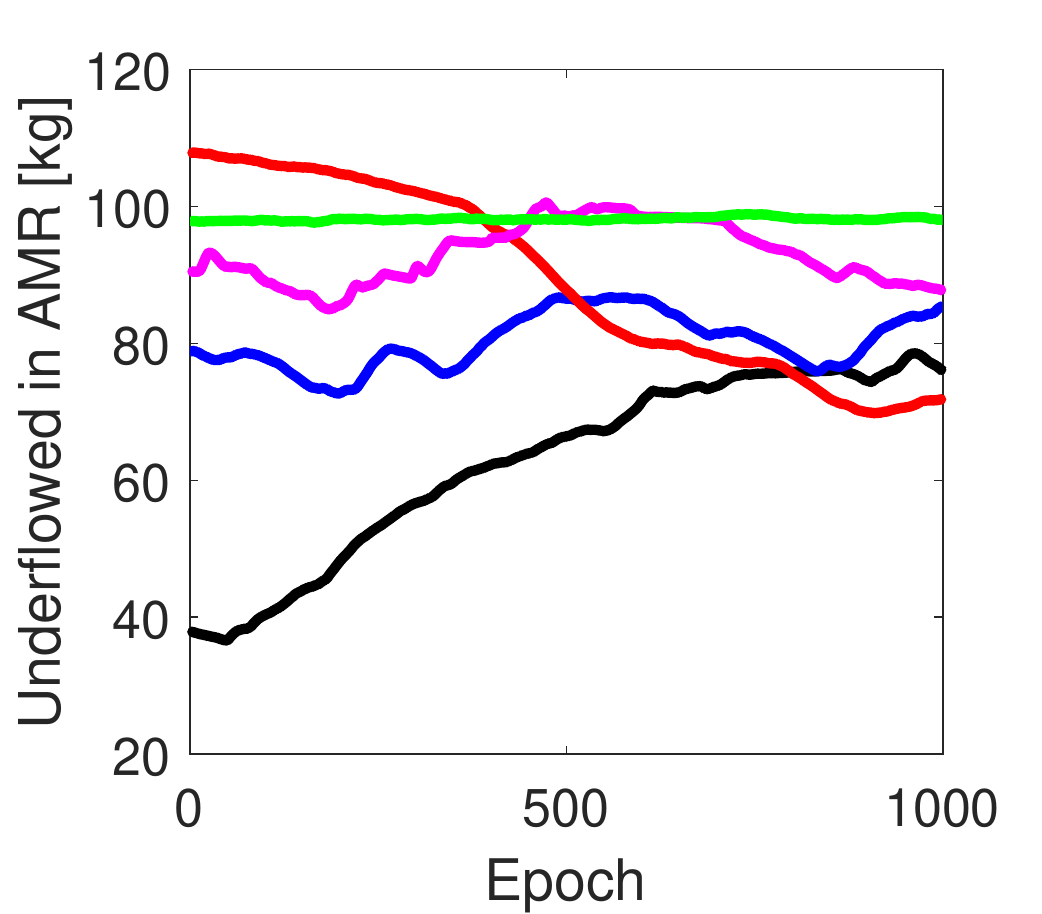} &
     \includegraphics[width=\linewidth]{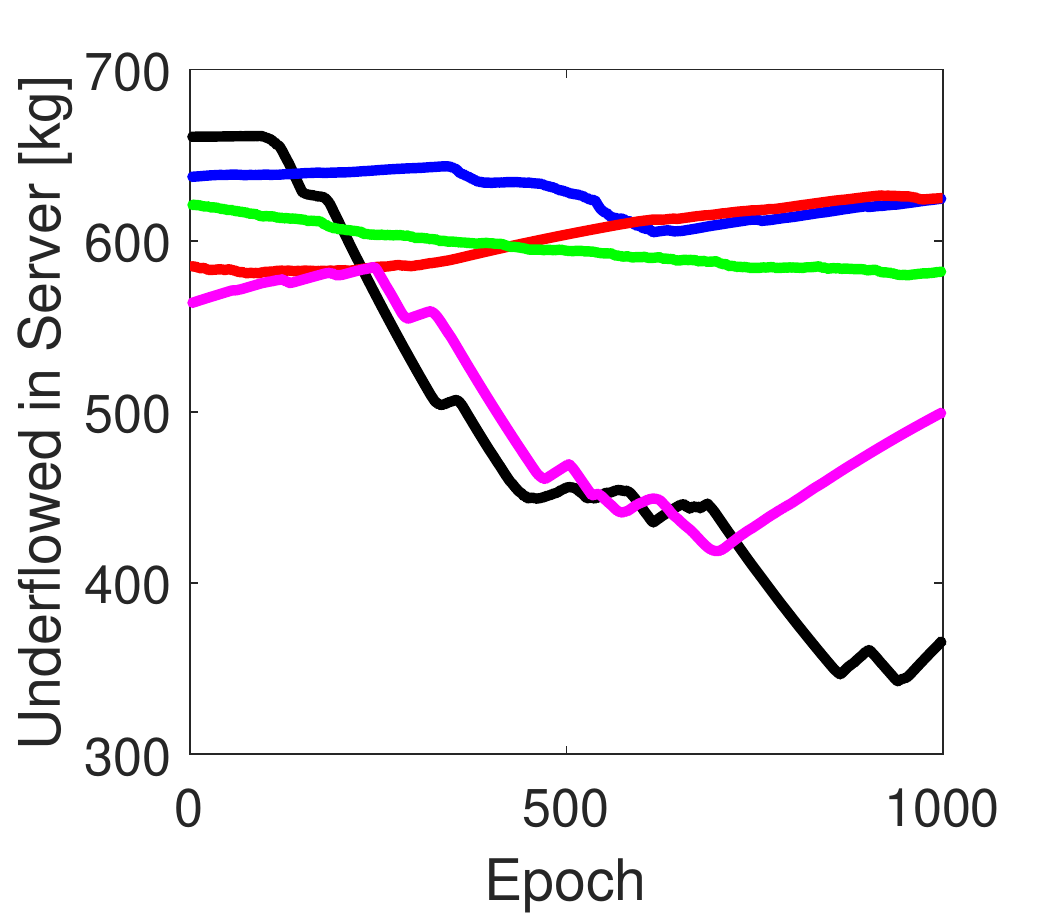} \\
     \multicolumn{1}{c}{\scriptsize(f) Overflowed load in AMR} & 
     \multicolumn{1}{c}{\scriptsize(g) Overflowed load in warehouse}   & 
     \multicolumn{1}{c}{\scriptsize(h) Underflowed load in AMR}  & 
     \multicolumn{1}{c}{\scriptsize(i) Underflowed load in warehouse}  
\end{tabular}
     \caption{The experimental result of various metrics with comparing different MARL frameworks.}
\label{fig:4}
\end{figure*}  

\begin{table*}[t!]
    \centering

\normalsize
    \caption{The numerical results of various metrics corresponding to the different benchmark schemes.}\begin{tabular}{c||ccccc}
        \toprule[1pt]
        \textbf{Metric}&\multicolumn{5}{c}{\bf {Benchmark Scheme}} \\ 
        \textbf{[SI Unit of (a)--(f): kg]}& \textsf{Proposed} & \textsf{Comp1} & \textsf{Comp2} & \textsf{Comp3} & \textsf{Comp4} \\\midrule
        \multicolumn{1}{l||}{(a) Avg. load status of AMR}&  6.0 &  2.9 &  9.3 & 2.9  & 2.6\\
        \multicolumn{1}{l||}{(b) Avg. load status of server }& 511 & 
 88 & 87 & 244 & 87\\
        \multicolumn{1}{l||}{(c) Avg. overflowed load in AMR}& \textbf{81} & 101& 224& 131& 103\\
        \multicolumn{1}{l||}{(d) Avg. overflowed load in server}& \textbf{2.6} & 0 & 0 & 5.1  & 0\\
        \multicolumn{1}{l||}{(e) Avg. underflowed load in AMR} &  \textbf{77} &  106&  227&  136 & 100\\
        \multicolumn{1}{l||}{(f) Avg. underflowed load in server}& \textbf{371} & 628 & 630 & 493  & 579\\\midrule
        \multicolumn{1}{l||}{(g) Avg. precision of load [\%]}& \textbf{92.1\%} & 90.8\% & 90.8\%& 92.2\% & 89.3\%\\
        \multicolumn{1}{l||}{(h) Avg. processing time [Minute]}& \textbf{292} & 294 & 255& 371 & 253\\
        \bottomrule[1pt]
    \end{tabular}
    \label{tab:4}
\end{table*}

\subsection{Experimental Setup}\label{sec:5A}
To verify the effectiveness of the proposed QMARL framework for smart factory management (named, \textsf{Proposed}), the proposed QMARL-based algorithm is compared with four comparing methods as listed in Table~\ref{tab:3}. 
The purpose of these numerical experiments is as follows,
\begin{itemize}
    \item The comparative experiments of \textsf{Proposed}, \textsf{Comp1}, and \textsf{Comp2} are conducted to corroborate the quantum advantages. The number of parameters is equally set for a fair comparison.
    \item This paper compares \textsf{Proposed} and \textsf{Comp3} to verify that the proposed method can achieve better performance than the latest MARL technique.
    \item To verify the superiority of MARL, this paper compares MARL schemes to random walk schemes, i.e., \textsf{Comp4}.
    \item To investigate the robustness of quality control, we train benchmark schemes in various environments regarding precision. We validate the fact that the quality of the load is time-varying in the environment. We corroborate the robustness of the quality control in our proposed scheme.
\end{itemize}

The simulation parameter settings are listed in Table~\ref{tab:2}. Because the number of qubits used in this paper is lower than $110$, this paper assumes that quantum noise is negligible.
\textsf{Comp1} is a hybrid quantum classical method utilizing A2C critic structure which is proposed and developed in another work~\cite{schenk2022hybrid}. 
Moreover, \textsf{Comp2} and \textsf{Comp3} are based on CTDE structure. Specifically, the value decomposition network (VDN)~\cite{sunehag2017value}.
For a fair comparison, we compose the neural network of linear operations and activation functions (i.e., linear or dense layer).
The python software libraries (\texttt{torchquantum} and \texttt{pytorch}) are used for deploying VQCs and DL methods, which support GPU acceleration~\cite{hanruiwang2022quantumnas}. %The baseline framework supporting this study's findings is available in \cite{yun2022quantum}. 
In addition, all experiments are conducted on a multi-GPU platform (equipped with 2 NVIDIA Titan XP GPUs using a 1405 MHz main clock and 12 GB memory) for training and inferencing/testing.

\subsection{Performance of Training} 
Fig.~\ref{fig:4} presents the numerical results corresponding to the training metric. This paper adopts total reward, precision, processing time, and loaded/overflowed/underflowed amount in AMR/server as training metrics.
As shown in Fig.~\ref{fig:4}(a), all training benchmark schemes (\textit{i.e.,} \textsf{Proposed}, \textsf{Comp1}, \textsf{Comp2} and \textsf{Comp3}) converge to the expected value for each different total reward. \textsf{Proposed}, which utilizes VQCs for both actor and critic network configuration, can observe the increased total reward from the beginning of learning to $980$ epochs. Then, the total reward of \textsf{Proposed} achieves the final value of $-37$. 
\textsf{Comp1} and \textsf{Comp2}, which share a common state-value network composed of a small number of parameters, do not evaluate their values properly and cause the reward to exist between $-205$ and $-240$. This is lower than $-200$, which is the expected value of the total reward when a random walk is performed. However, a classical actor and critic composed of a large number of parameters $(\approx40$K$)$ show similar performance to the proposed scheme (\textit{e.g.,} $\pm 10$ performance difference in the total reward).

In \textsf{Proposed} and \textsf{Comp3}, policy evaluation and improvement are trained to increase the total reward. However, \textsf{Comp1} and \textsf{Comp2} with classical critic networks composed of small parameters are trained (i.e., actor loss and critic loss are reduced), but not in the direction of reward increasing. In other words, policy evaluation and improvement are not working correctly.
The only difference between \textsf{Proposed} and \textsf{Comp1} is whether the critic is a quantum-based or a classical critic, and there is a huge difference in training performance. In addition, compared with \textsf{Comp3}, the number of parameters is $364\mathsf{x}$ lower than that of \textsf{Comp3}, whereas the performances are almost equivalent to each other.

\subsection{Feasibility Studies in the LCD Smart Factory Environment} 
This section investigates the proposed model's performance in LCD smart factory environment. Fig.~\ref{fig:4} shows the results of various metrics in the training process, and Table~\ref{tab:4} shows the performance after training is finished. The result of Table~\ref{tab:4} represents the average value of inference of $100$ iterations. 
Fig.~\ref{fig:4}(b--c) represent the average quality and processing time, respectively. 
Fig.~\ref{fig:4}(d--i) show the amount of loaded, overflowed, and underflowed load amount generated in warehouse and AMR, respectively. For this simulation, the total amount of overflowed/underflowed loads achieve target performance if the corresponding values become 0.
During the training process, it is shown that the values of the two metrics (i.e., the amount of overflowed/underflowed loads) of the \textsf{Proposed} and \textsf{Comp3} are reduced. Therefore, it can be inferred that \textsf{Proposed} and \textsf{Comp3} are trained in the correct direction. On the other hand, overflowed loads sparsely occurred in \textsf{Comp1} and \textsf{Comp2}. However, the underflowed load amount is the highest, which proves that \textsf{Comp1} and \textsf{Comp2} are learning in the direction that satisfies only one of the two goals. Furthermore, this tendency also affects the average load status of the warehouse and AMR agents. In the proposed scheme of this paper, it is confirmed that all four values of the indicators continuously decrease until they reach the approximate value of 0. Hence, it is confirmed that the AMR agent has the average load of 3.6\,kg in Fig.~\ref{fig:4}.

\subsection{Impact on State Encoding Method}
\begin{figure}[t!]
    \centering    \includegraphics[width=.79\columnwidth]{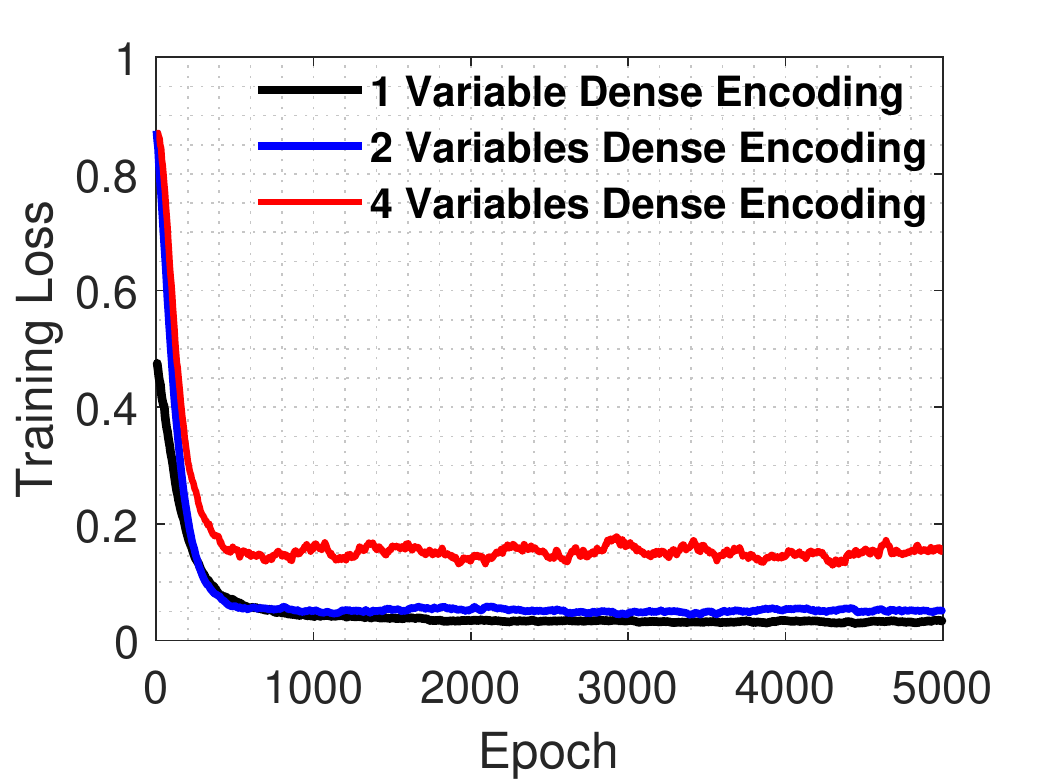}
    \caption{The mini-experiment for benchmarking various state encoding methods.}
    \label{fig:6}
\end{figure}

According to \cite{lockwood21, lockwood2020reinforcement}, state-encoding is crucial for the performance of QRL. Therefore, an experiment is designed to demonstrate the importance of state encoding. 
This experiment aims to transform four random bits into continuous scalar values. The output value is calculated as $ y = \sum_{i=1}^4{x_i} * 2^{1-i}$. In this transformation process, $\{4,2,1\}$ variables dense encoding is carried out to compare the dense encoding methods. Note that the number of parameters in VQCs is identical to 50. The result is shown in Fig. \ref{fig:6}, and it is concluded that the performance of $1,2$ variables dense encoding is high while the performance of $4$ variables dense encoding is low. In other words, the $2$-variables encoding used in this paper has less performance degradation than the $4$-variables encoding technique.

\begin{figure}[t!]
    \centering
    \includegraphics[width=.85\columnwidth]{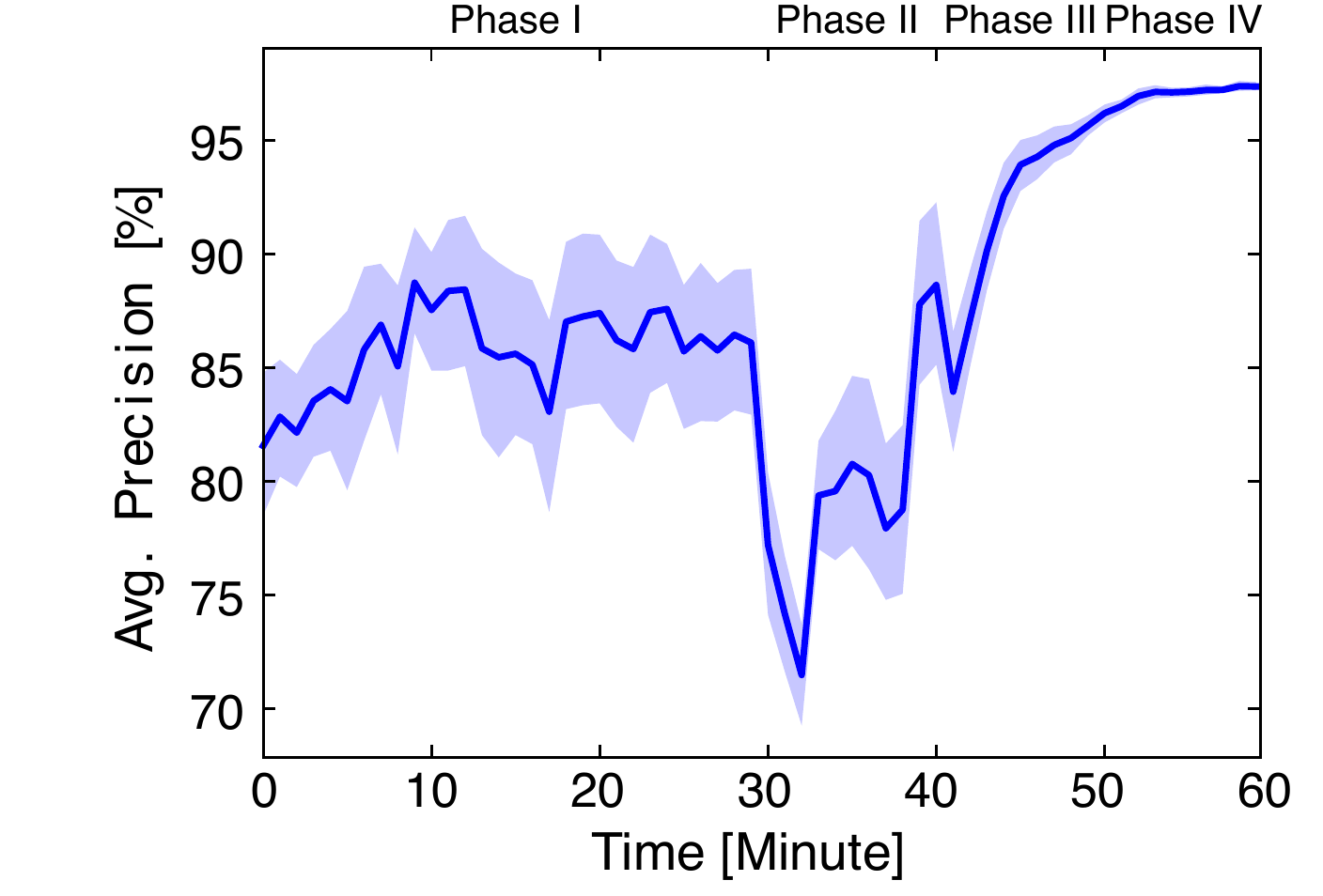}
    \caption{The robustness of quality control with time-varying quality of loads. }
    \label{fig:5}
\end{figure}

\subsection{Robustness of Quality Control}
We design the experiment to investigate the robustness of the proposed framework. To benchmark the robustness, we design the smart factory environment that is time-varying and configure the environment in four phases. In phase 1, the precision of load is randomly selected from $\{61.9, 95.8, 97.1\}\%$, which is identical to the training environment. Note that the initial precision follows the uniform distribution $\mathcal{U}[61.9, 97.1]\%$. The quality of LCD load carried by each AMR varies with time (e.g., $61.9$\%, $95.8$\%, and $97.1$\% for phase 2, phase 3, and phase 4, respectively). Then, the average precision is measured for 60 minutes to investigate the robustness of quality control.
The result of Fig.~\ref{fig:5} represents the average precision value of inference of $100$ iterations.
During phase 1, the precision records $88.4\%$ on average. At $t=30$, the quality of the input load decreases, i.e., the input load's precision equals $61.9\%$. Thus, the precision is $72.4\%$, the lowest precision during the episode. In response to this result, the AMR agents try to improve the precision value during phase 2. On the other hand, the AMR agents do not make actions on quality control in phases 3 and 4 since the quality of input load increases from $61.9\%$ to $97.1\%$.
In summary, the robustness of quality control in our proposed scheme is corroborated by demonstrating the ability of our AMR agents to encounter and cope with the unpredictable quality of input load.

\subsection{Discussion}
%\subsubsection{Impact on the Number of Parameters} 
This section provides in-depth discussions to explain why the proposed scheme outperforms the other frameworks. 
\subsubsection{Expressibility of Trainable Parameters}
The authors of \cite{abbas2021power} have argued that the parameters of VQCs have more expressibility for quantum neural networks than classical neural networks. The small number of trainable parameters in the RL/MARL regime acts as a vulnerability for the classical neural network. In \cite{lockwood2020reinforcement,yun2022quantum}, it is proven that QRL and QMARL can achieve similar performance to a classical RL/MARL. In the results of this paper, the classical neural network with fewer parameters yields lower performance due to two reasons; 1) index embedding on observation, and 2) parameter-shared policy. The index embedding on agents' observations and the parameter-shared policy method are utilized for faster convergence despite taking a loss in performance.
Furthermore, the expressibility capacity of a neural network is also trusted to be sufficient, which is why the two methods introduced above are used~\cite{ICML2018_QMIX}. Unfortunately, the degradation in performance is significant regardless of the expressibility capacity. On the other hand, the quantum circuit operates successfully even with a small number of parameters. 
\subsubsection{Dimensional Reduction Corresponding to the State Encoding}
Information loss occurs when the input variables are lost by dimensional reduction. In the experiments, the dimension of the input variable is set to four, and the output variable is set to four, two, and one for different schemes, respectively. In four variables dense encoding, the information loss is severe, because four independent variables are encoded using four rotation gates $R_Y(x_4)$, $R_Y(x_3)$, $R_Z(x_2)$, and $R_Z(x_1)$ through a single qubit. In the cases of two variables dense encoding and one variable dense encoding, the dimensional reduction does not occur. This is proven by showing the encoding processes on the Bloch sphere. For two and one variables encoding, the qubits are rotated twice in two orthogonal directions (e.g., $y$-axis and $z$-axis directions) and once in one direction, respectively. Consequently, the ranks of the resultant qubits are guaranteed. Therefore, the four variables dense encoding method has the lowest performance and is outperformed by the other aforementioned methods.

\section{Concluding Remarks}\label{sec:6}
This work has investigated the design of QMARL agents based on VQCs for autonomous multi-robot control and coordination in smart factory management while taking POMDP into consideration. When utilizing AMRs as QMARL agents, the two variables dense encoding method is implemented to reduce the number of qubits in the proposed model. In addition, this paper adopts the parameter-shared policy with index embedding, which can reduce the number of trainable parameters. Using the abovementioned techniques, the quantum policy and state-value function are configured to quantum multi-agent actor-critic. The extensive numerical results show the superiority of the proposed QMARL-based AMR control in smart factory management. Finally, the proposed QMARL has an explicit performance gain when using the same number of parameters compared to the classical MARL algorithm and does not suffer from a severe dimensional reduction of data compared to other state-encoding methods.
\bibliographystyle{IEEEtran}  
\bibliography{ref_aimlab,ref_quantum,ref_rl}

\begin{thebibliography}{10}
\providecommand{\url}[1]{#1}
\csname url@samestyle\endcsname
\providecommand{\newblock}{\relax}
\providecommand{\bibinfo}[2]{#2}
\providecommand{\BIBentrySTDinterwordspacing}{\spaceskip=0pt\relax}
\providecommand{\BIBentryALTinterwordstretchfactor}{4}
\providecommand{\BIBentryALTinterwordspacing}{\spaceskip=\fontdimen2\font plus
\BIBentryALTinterwordstretchfactor\fontdimen3\font minus
  \fontdimen4\font\relax}
\providecommand{\BIBforeignlanguage}[2]{{%
\expandafter\ifx\csname l@#1\endcsname\relax
\typeout{** WARNING: IEEEtran.bst: No hyphenation pattern has been}%
\typeout{** loaded for the language `#1'. Using the pattern for}%
\typeout{** the default language instead.}%
\else
\language=\csname l@#1\endcsname
\fi
#2}}
\providecommand{\BIBdecl}{\relax}
\BIBdecl

\bibitem{icdcs22yun}
W.~J. Yun, Y.~Kwak, J.~P. Kim, H.~Cho, S.~Jung, J.~Park, and J.~Kim, ``Quantum
  multi-agent reinforcement learning via variational quantum circuit design,''
  in \emph{Proceedings of the IEEE International Conference on Distributed
  Computing Systems (ICDCS)}, Bologna, Italy, July 2022.

\bibitem{iot103}
W.~Na, Y.~Lee, N.-N. Dao, D.~N. Vu, A.~Masood, and S.~Cho, ``Directional link
  scheduling for real-time data processing in smart manufacturing system,''
  \emph{IEEE Internet of Things Journal}, vol.~5, no.~5, pp. 3661--3671, 2018.

\bibitem{8310596}
S.~Jeong, W.~Na, J.~Kim, and S.~Cho, ``Internet of things for smart
  manufacturing system: Trust issues in resource allocation,'' \emph{IEEE
  Internet of Things Journal}, vol.~5, no.~6, pp. 4418--4427, December 2018.

\bibitem{iot101}
W.~Yu, Y.~Liu, T.~Dillon, W.~Rahayu, and F.~Mostafa, ``An integrated framework
  for health state monitoring in a smart factory employing {IoT} and big data
  techniques,'' \emph{IEEE Internet of Things Journal}, vol.~9, no.~3, pp.
  2443--2454, 2022.

\bibitem{iot102}
S.~Savazzi, V.~Rampa, and U.~Spagnolini, ``Wireless cloud networks for the
  factory of things: Connectivity modeling and layout design,'' \emph{IEEE
  Internet of Things Journal}, vol.~1, no.~2, pp. 180--195, 2014.

\bibitem{9699416}
Y.~Jin, B.~Huang, Y.~Yan, Y.~Huan, J.~Xu, S.~Li, P.~Gope, L.~Xu, Z.~Zou, and
  L.~Zheng, ``Edge-based collaborative training system for artificial
  intelligence-of-things,'' \emph{IEEE Transactions on Industrial Informatics},
  pp. 1--1, 2022.

\bibitem{9650739}
Z.~Nie and K.-C. Chen, ``Hypergraphical real-time multi-robot task allocation
  in a smart factory,'' \emph{IEEE Transactions on Industrial Informatics}, pp.
  1--1, 2021.

\bibitem{8950429}
G.~Fortino, F.~Messina, D.~Rosaci, G.~M.~L. Sarné, and C.~Savaglio, ``A
  trust-based team formation framework for mobile intelligence in smart
  factories,'' \emph{IEEE Transactions on Industrial Informatics}, vol.~16,
  no.~9, pp. 6133--6142, September 2020.

\bibitem{iot104}
S.~Jeong, W.~Na, J.~Kim, and S.~Cho, ``Internet of things for smart
  manufacturing system: Trust issues in resource allocation,'' \emph{IEEE
  Internet of Things Journal}, vol.~5, no.~6, pp. 4418--4427, 2018.

\bibitem{9247159}
E.~A. Oyekanlu, A.~C. Smith, W.~P. Thomas, G.~Mulroy, D.~Hitesh, M.~Ramsey,
  D.~J. Kuhn, J.~D. Mcghinnis, S.~C. Buonavita, N.~A. Looper, M.~Ng,
  A.~Ng'oma, W.~Liu, P.~G. Mcbride, M.~G. Shultz, C.~Cerasi, and D.~Sun, ``A
  review of recent advances in automated guided vehicle technologies:
  Integration challenges and research areas for {5G}-based smart manufacturing
  applications,'' \emph{IEEE Access}, vol.~8, pp. 202\,312--202\,353, 2020.

\bibitem{9216961}
W.~Xia, J.~Goh, C.~A. Cortes, Y.~Lu, and X.~Xu, ``Decentralized coordination of
  autonomous {AGVs} for flexible factory automation in the context of {Industry
  4.0},'' in \emph{Proc. IEEE Int'l Conf. on Automation Science and Engineering
  (CASE)}, 2020, pp. 488--493.

\bibitem{verizon}
\BIBentryALTinterwordspacing
P.~Apte, ``What are autonomous mobile robots, and how can they transform
  manufacturing?'' \emph{Verizon Business Resources and Industry Insights
  Articles}, October 2021. [Online]. Available:
  \url{https://www.verizon.com/business/resources/articles/s/autonomous-mobile-robots-can-drive-smart-manufacturing/}
\BIBentrySTDinterwordspacing

\bibitem{hu2020voronoi}
J.~Hu, H.~Niu, J.~Carrasco, B.~Lennox, and F.~Arvin, ``Voronoi-based
  multi-robot autonomous exploration in unknown environments via deep
  reinforcement learning,'' \emph{IEEE Transactions on Vehicular Technology},
  vol.~69, no.~12, pp. 14\,413--14\,423, December 2020.

\bibitem{9682599}
W.~J. Yun, S.~Park, J.~Kim, M.~Shin, S.~Jung, A.~Mohaisen, and J.-H. Kim,
  ``Cooperative multi-agent deep reinforcement learning for reliable
  surveillance via autonomous multi-{UAV} control,'' \emph{IEEE Transactions on
  Industrial Informatics}, pp. 1--1, 2022.

\bibitem{schuld2022quantum}
M.~Schuld and N.~Killoran, ``Is quantum advantage the right goal for quantum
  machine learning?'' \emph{CoRR}, vol. abs:2203.01340, 2022.

\bibitem{hanruiwang2022quantumnas}
H.~Wang, Y.~Ding, J.~Gu, Z.~Li, Y.~Lin, D.~Z. Pan, F.~T. Chong, and S.~Han,
  ``Quantum{NAS}: Noise-adaptive search for robust quantum circuits,'' in
  \emph{Proc. IEEE Int'l Symposium on High-Performance Computer Architecture
  (HPCA)}, April 2022.

\bibitem{yun2022pvm}
W.~J. Yun, H.~Baek, and J.~Kim, ``Projection valued measure-based quantum
  machine learning for multi-class classification,'' \emph{CoRR}, vol.
  abs/2210.16731, 2022.

\bibitem{Cornelissen22STOC}
A.~Cornelissen, Y.~Hamoudi, and S.~Jerbi, ``Near-optimal quantum algorithms for
  multivariate mean estimation,'' in \emph{Proc. of ACM SIGACT Symposium on
  Theory of Computing (STOC)}, New York, NY, USA, 2022, p. 33–43.

\bibitem{wiedemann2022quantum}
S.~Wiedemann, D.~Hein, S.~Udluft, and C.~Mendl, ``Quantum policy iteration via
  amplitude estimation and grover search--towards quantum advantage for
  reinforcement learning,'' \emph{arXiv preprint arXiv:2206.04741}, 2022.

\bibitem{arute2019quantum}
F.~Arute, K.~Arya, R.~Babbush, D.~Bacon, J.~C. Bardin, R.~Barends, R.~Biswas,
  S.~Boixo, F.~G. Brandao, D.~A. Buell \emph{et~al.}, ``Quantum supremacy using
  a programmable superconducting processor,'' \emph{Nature}, vol. 574, no.
  7779, pp. 505--510, 2019.

\bibitem{oh2020qcnn-simple}
S.~Oh, J.~Choi, and J.~Kim, ``A tutorial on quantum convolutional neural
  networks {(QCNN)},'' in \emph{Proc. IEEE Int'l Conf. on ICT Convergence
  (ICTC)}, October 2020.

\bibitem{jerbi2021variational}
S.~Jerbi, C.~Gyurik, S.~Marshall, H.~J. Briegel, and V.~Dunjko, ``Variational
  quantum policies for reinforcement learning,'' in \emph{Proc. Neural
  Information Processing Systems (NeurIPS)}, December 2021.

\bibitem{ijcnn21hong}
Z.~Hong, J.~Wang, X.~Qu, X.~Zhu, J.~Liu, and J.~Xiao, ``Quantum convolutional
  neural network on protein distance prediction,'' in \emph{Proc. IEEE Int'l
  Joint Conf. on Neural Networks (IJCNN)}, July 2021.

\bibitem{icufn21kwak}
Y.~Kwak, W.~J. Yun, S.~Jung, and J.~Kim, ``Quantum neural networks: Concepts,
  applications, and challenges,'' in \emph{Proc. IEEE Int'l Conf. on Ubiquitous
  and Future Networks (ICUFN)}, August 2021.

\bibitem{chen2020variational}
S.~Y.-C. Chen, C.-H.~H. Yang, J.~Qi, P.-Y. Chen, X.~Ma, and H.-S. Goan,
  ``Variational quantum circuits for deep reinforcement learning,'' \emph{IEEE
  Access}, vol.~8, pp. 141\,007--141\,024, 2020.

\bibitem{ictc21kwak}
Y.~Kwak, W.~J. Yun, S.~Jung, J.-K. Kim, and J.~Kim, ``Introduction to quantum
  reinforcement learning: Theory and {PennyLane}-based implementation,'' in
  \emph{Proc. IEEE Int'l Conf. on ICT Convergence (ICTC)}, October 2021.

\bibitem{carleo2019machine}
G.~Carleo, I.~Cirac, K.~Cranmer, L.~Daudet, M.~Schuld, N.~Tishby,
  L.~Vogt-Maranto, and L.~Zdeborov{\'a}, ``Machine learning and the physical
  sciences,'' \emph{Reviews of Modern Physics}, vol.~91, no.~4, p. 045002,
  2019.

\bibitem{arxiv2017_VDN}
P.~Sunehag, G.~Lever, A.~Gruslys, W.~M. Czarnecki, V.~Zambaldi, M.~Jaderberg,
  M.~Lanctot, N.~Sonnerat, J.~Z. Leibo, K.~Tuyls \emph{et~al.},
  ``Value-decomposition networks for cooperative multi-agent learning,''
  \emph{CoRR}, June 2017.

\bibitem{shor1995scheme}
P.~W. Shor, ``Scheme for reducing decoherence in quantum computer memory,''
  \emph{Physical Review A}, vol.~52, no.~4, p. R2493, 1995.

\bibitem{yun2022quantum}
W.~J. Yun, Y.~Kwak, J.~P. Kim, H.~Cho, S.~Jung, J.~Park, and J.~Kim, ``Quantum
  multi-agent reinforcement learning via variational quantum circuit design,''
  \emph{arXiv preprint arXiv:2203.10443}, 2022.

\bibitem{lockwood2020reinforcement}
O.~Lockwood and M.~Si, ``Reinforcement learning with quantum variational
  circuit,'' in \emph{Proc. AAAI Conference on Artificial Intelligence and
  Interactive Digital Entertainment (AIIDE)}, October 2020.

\bibitem{simeone2022introduction}
O.~Simeone, ``An introduction to quantum machine learning for engineers,''
  \emph{CoRR}, vol. abs/2205.09510, June 2022.

\bibitem{Shor94}
P.~W. Shor, ``Algorithms for quantum computation: Discrete logarithms and
  factoring,'' in \emph{Proc. IEEE Foundations of Computer Science (FOCS)},
  Santa Fe, NM, USA, November 1994, pp. 124--134.

\bibitem{Grover96STOC}
L.~K. Grover, ``A fast quantum mechanical algorithm for database search,'' in
  \emph{Proc. of ACM Symposium on Theory of Computing (STOC)}, ser. STOC '96,
  New York, NY, USA, 1996, p. 212–219.

\bibitem{8988635}
A.~Bolu and Ã.~Korçak, ``Path planning for multiple mobile robots in smart
  warehouse,'' in \emph{Proc. IEEE International Conference on Control,
  Mechatronics and Automation (ICCMA)}, 2019, pp. 144--150.

\bibitem{villalonga2020establishing}
B.~Villalonga, D.~Lyakh, S.~Boixo, H.~Neven, T.~S. Humble, R.~Biswas, E.~G.
  Rieffel, A.~Ho, and S.~Mandr{\`a}, ``Establishing the quantum supremacy
  frontier with a 281 pflop/s simulation,'' \emph{Quantum Science and
  Technology}, vol.~5, no.~3, p. 034003, 2020.

\bibitem{LCD-REFs}
X.~Bai, Y.~Fang, W.~Lin, L.~Wang, and B.-F. Ju, ``Saliency-based defect
  detection in industrial images by using phase spectrum,'' \emph{IEEE
  Transactions on Industrial Informatics}, vol.~10, no.~4, pp. 2135--2145,
  2014.

\bibitem{lee2017robust}
J.-Y. Lee, T.-W. Kim, and H.~J. Pahk, ``Robust defect detection method for a
  non-periodic tft-lcd pad area,'' \emph{International Journal of Precision
  Engineering and Manufacturing}, vol.~18, no.~8, pp. 1093--1102, 2017.

\bibitem{LCDTFT}
Y.~Xia, C.~Luo, Y.~Zhou, and L.~Jia, ``A hybrid method of frequency and spatial
  domain techniques for {TFT-LCD} circuits defect detection,'' \emph{IEEE
  Transactions on Semiconductor Manufacturing}, pp. 1--1, 2022.

\bibitem{xue2017trajectory}
T.~Xue, R.~Li, M.~Tokgo, J.~Ri, and G.~Han, ``Trajectory planning for
  autonomous mobile robot using a hybrid improved qpso algorithm,'' \emph{Soft
  Computing}, vol.~21, no.~9, pp. 2421--2437, 2017.

\bibitem{tvt202106jung}
S.~Jung, W.~J. Yun, M.~Shin, J.~Kim, and J.-H. Kim, ``Orchestrated scheduling
  and multi-agent deep reinforcement learning for cloud-assisted multi-{UAV}
  charging systems,'' \emph{IEEE Transactions on Vehicular Technology},
  vol.~70, no.~6, pp. 5362--5377, June 2021.

\bibitem{Springer2016_POMDP}
F.~A. Oliehoek and C.~Amato, \emph{A Concise Introduction to Decentralized
  {POMDPs}}.\hskip 1em plus 0.5em minus 0.4em\relax Springer Publishing
  Company, Incorporated, 2016.

\bibitem{crook19}
G.~E. Crooks, ``Gradients of parameterized quantum gates using the
  parameter-shift rule and gate decomposition,'' \emph{CoRR}, vol.
  abs/1905.13311, May 2019.

\bibitem{LCDweight}
\BIBentryALTinterwordspacing
D.~Company, ``Dell e series e2311h monitor,'' \emph{DELL Inc.}, 2022. [Online].
  Available:
  \url{http://www1.la.dell.com/content/products/productdetails.aspx/monitor-dell-e2311h?c=sr&l=en&s=corp&~tab=specstab}
\BIBentrySTDinterwordspacing

\bibitem{schenk2022hybrid}
M.~Schenk, E.~F. Combarro, M.~Grossi, V.~Kain, K.~S.~B. Li, and S.~Popa,
  Mircea-Marian aFnd~Vallecorsa, ``Hybrid actor-critic algorithm for quantum
  reinforcement learning at cern beam lines,'' \emph{CoRR}, vol.
  abs/2209.11044, September 2022.

\bibitem{sunehag2017value}
P.~Sunehag, G.~Lever, A.~Gruslys, W.~M. Czarnecki, V.~Zambaldi, M.~Jaderberg,
  M.~Lanctot, N.~Sonnerat, J.~Z. Leibo, K.~Tuyls \emph{et~al.},
  ``Value-decomposition networks for cooperative multi-agent learning,''
  \emph{CoRR}, vol. abs/1706.05296, June 2017.

\bibitem{lockwood21}
O.~Lockwood and M.~Si, ``Playing {Atari} with hybrid quantum-classical
  reinforcement learning,'' in \emph{Proc. NeurIPS 2020 Workshop on
  Pre-registration in Machine Learning}, December 2021, pp. 285--301.

\bibitem{abbas2021power}
A.~Abbas, D.~Sutter, C.~Zoufal, A.~Lucchi, A.~Figalli, and S.~Woerner, ``The
  power of quantum neural networks,'' \emph{Nature Computational Science},
  vol.~1, no.~6, pp. 403--409, 2021.

\bibitem{ICML2018_QMIX}
T.~Rashid, M.~Samvelyan, C.~Schroeder, G.~Farquhar, J.~Foerster, and
  S.~Whiteson, ``{QMIX}: Monotonic value function factorisation for deep
  multi-agent reinforcement learning,'' in \emph{Proc. of the International
  Conference on Machine Learning (ICML)}, Stockholmsm{\"a}ssan, Sweden, July
  2018, pp. 4295--4304.

\end{thebibliography}

\begin{IEEEbiography}[{\includegraphics[width=1in,height=1.25in,clip]{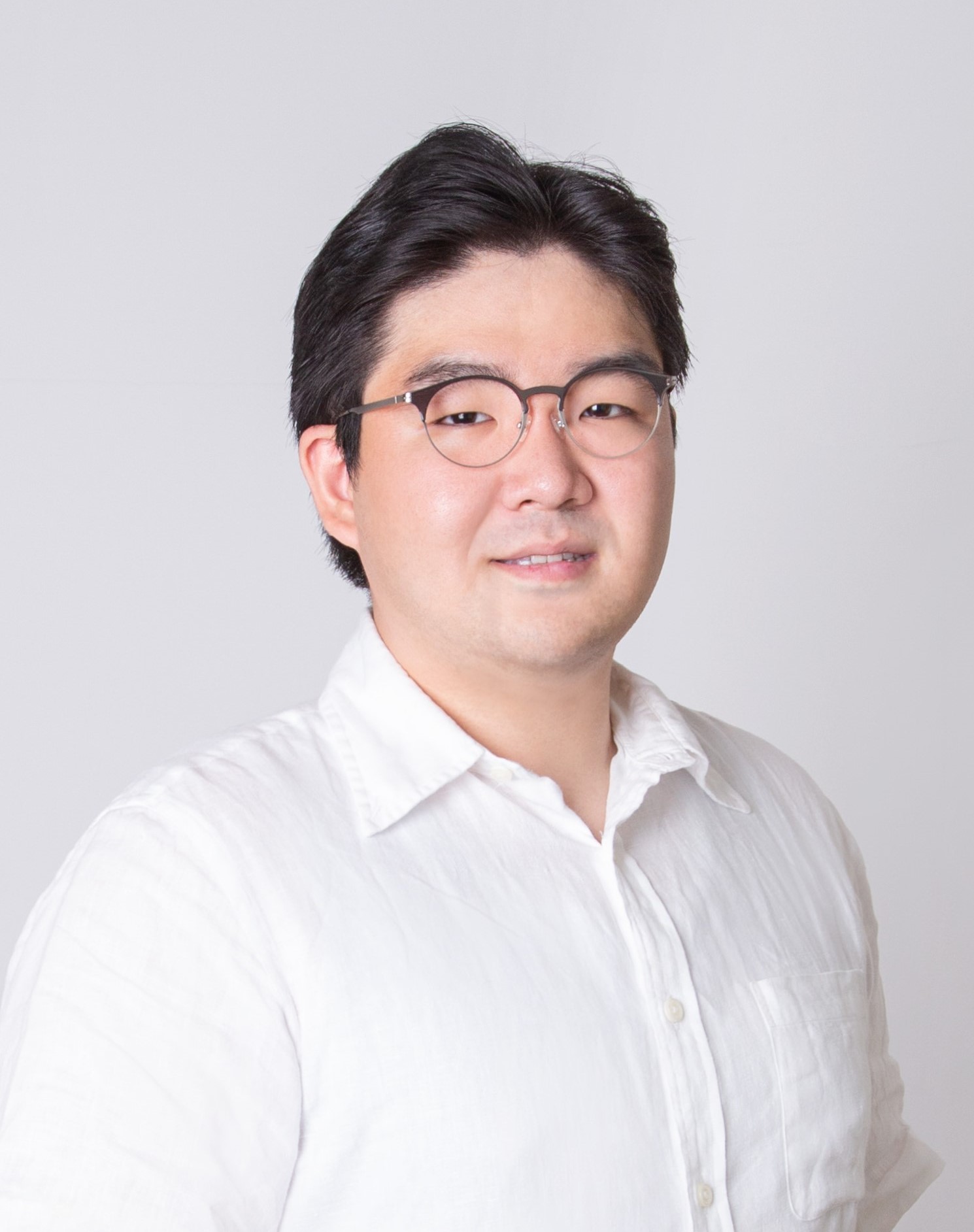}}]{Won Joon Yun} 
 is currently a Ph.D. student in electrical and computer engineering at Korea University, Seoul, Republic of Korea, since March 2021, where he received his B.S. in electrical engineering. He was a visiting researcher at Cipherome Inc., San Jose, CA, USA, during the summer of 2022; and also a visiting researcher at the University of Southern California, Los Angeles, CA, USA during the winter of 2022 for a joint project with Prof. Andreas F. Molisch at the Ming Hsieh Department of Electrical and Computer Engineering, USC Viterbi School of Engineering. 
His current research interests include machine learning in various fields, quantum machine learning, and multi-agent reinforcement learning. 
\end{IEEEbiography}

\begin{IEEEbiography}[{\includegraphics[width=1in,height=1.25in,clip]{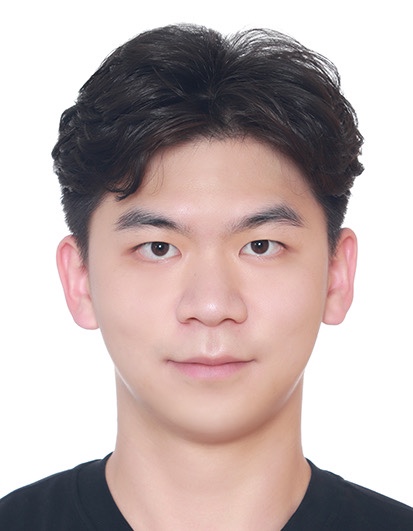}}]{Jae Pyoung Kim} is currently an M.S. student in electrical and computer engineering at Korea University, Seoul, Republic of Korea, since March 2023, where he received his B.S. in electrical engineering. student in electrical and computer engineering. 
He is a research engineer at the Artificial Intelligence and Mobility (AIM) Laboratory at Korea University, Seoul, Republic of Korea, from 2021 to 2022. 
His current research interests include quantum machine learning. 
\end{IEEEbiography}

\begin{IEEEbiography}[{\includegraphics[width=1in,height=1.25in,clip]{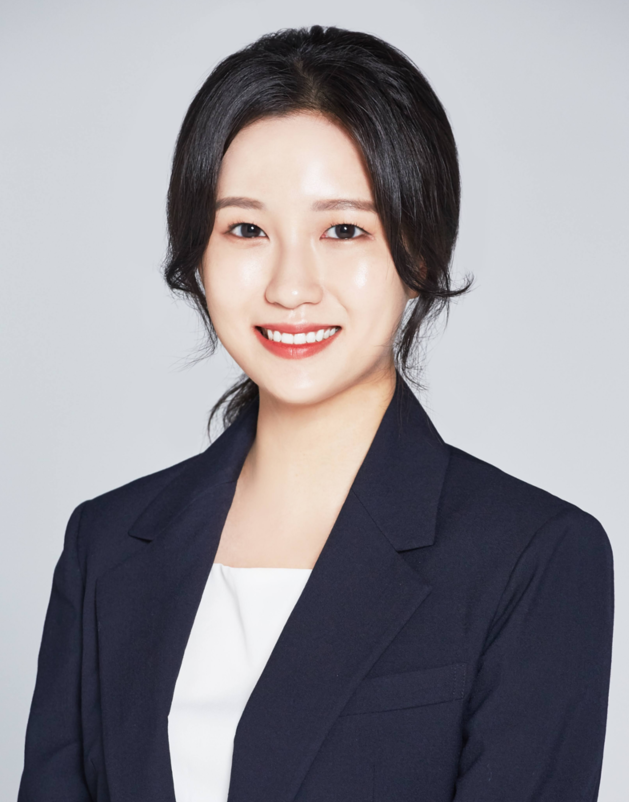}}]{Soyi Jung} has been an assistant professor at the department of electrical and computer engineering, Ajou University, Suwon, Republic of Korea, since September 2022. She also holds a visiting scholar position at Donald Bren School of Information and Computer Sciences, University of California, Irvine, CA, USA, from 2021 to 2022. She was a research professor at Korea University, Seoul, Republic of Korea, during 2021. She was also a researcher at Korea Testing and Research (KTR) Institute, Gwacheon, Republic of Korea, from 2015 to 2016. 
She received her B.S., M.S., and Ph.D. degrees in electrical and computer engineering from Ajou University, Suwon, Republic of Korea, in 2013, 2015, and 2021, respectively. 

Her current research interests include network optimization for autonomous vehicles communications, distributed system analysis, big-data processing platforms, and probabilistic access analysis. She was a recipient of Best Paper Award by KICS (2015), Young Women Researcher Award by WISET and KICS (2015), Bronze Paper Award from IEEE Seoul Section Student Paper Contest (2018), ICT Paper Contest Award by Electronic Times (2019), and IEEE ICOIN Best Paper Award (2021).
\end{IEEEbiography}

\begin{IEEEbiography}[{\includegraphics[width=1in,height=1.25in,clip]{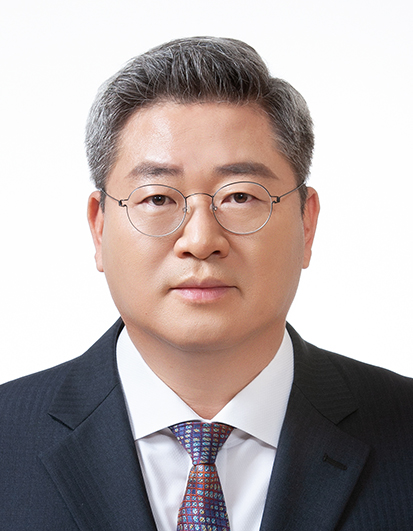}}]{Jae-Hyun Kim} received the B.S., M.S., and Ph.D. degrees, all in computer science and engineering, from Hanyang University, Ansan, Korea, in 1991, 1993, and 1996 respectively. In 1996, he was with the Communication Research Laboratory, Tokyo, Japan, as a Visiting Scholar. From April 1997 to October 1998, he was a postdoctoral fellow at the department of electrical engineering, University of California, Los Angeles. From November 1998 to February 2003, he worked as a member of technical staff in Performance Modeling and QoS management department, Bell laboratories, Lucent Technologies, Holmdel, NJ. He has been with the department of electrical and computer engineering, Ajou University, Suwon, Korea, as a professor since 2003. 

He is the Center Chief of Satellite Information Convergence Application Services Research Center (SICAS) sponsored by Institute for Information $\&$ Communications Technology Promotion in Korea. He is Chairman of the Smart City Committee of 5G Forum in Korea since 2018. He is vice president of the Korea Institute of Communication and Information Sciences (KICS) from 2022.  He is a member of the IEEE, KICS, the Institute of Electronics and Information Engineers (IEIE), and the Korean Institute of Information Scientists and Engineers (KIISE). He was a recipient of IEEE ICOIN Best Paper Award (2021).
\end{IEEEbiography}

\begin{IEEEbiography}[{\includegraphics[width=1in,height=1.25in,clip]{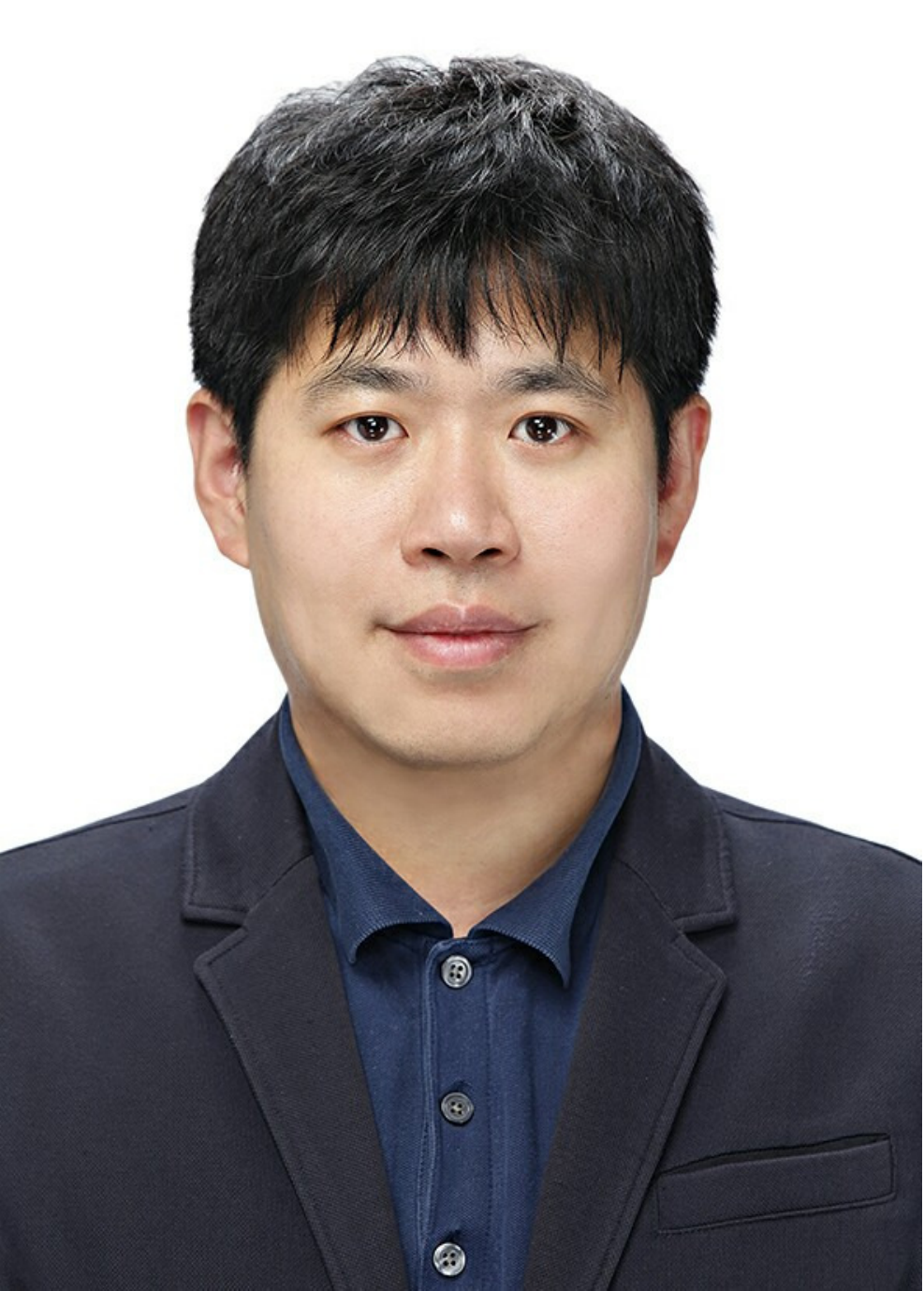}}]{Joongheon Kim}
(M'06--SM'18) has been with Korea University, Seoul, Korea, since 2019, where he is currently an associate professor at the School of Electrical Engineering and also an adjunct professor at the Department of Communications Engineering (established/sponsored by Samsung Electronics) and the Department of Semiconductor Engineering (established/sponsored by SK Hynix). He received the B.S. and M.S. degrees in computer science and engineering from Korea University, Seoul, Korea, in 2004 and 2006; and the Ph.D. degree in computer science from the University of Southern California (USC), Los Angeles, CA, USA, in 2014. Before joining Korea University, he was a research engineer with LG Electronics (Seoul, Korea, 2006--2009), a systems engineer with Intel Corporation (Santa Clara, CA, USA, 2013--2016), and an assistant professor of computer science and engineering with Chung-Ang University (Seoul, Korea, 2016--2019). 

He serves as an editor for \textsc{IEEE Transactions on Vehicular Technology}, \textsc{IEEE Transactions on Machine Learning in Communications and Networking}, and \textsc{IEEE Communications Standards Magazine}. He is also a distinguished lecturer for \textit{IEEE Communications Society (ComSoc)} and \textit{IEEE Systems Council}.

He was a recipient of Annenberg Graduate Fellowship with his Ph.D. admission from USC (2009), Intel Corporation Next Generation and Standards (NGS) Division Recognition Award (2015), \textsc{IEEE Systems Journal} Best Paper Award (2020), IEEE ComSoc Multimedia Communications Technical Committee (MMTC) Outstanding Young Researcher Award (2020), IEEE ComSoc MMTC Best Journal Paper Award (2021), and Best Special Issue Guest Editor Award by \textit{ICT Express (Elsevier)} (2022). He also received several awards from IEEE conferences including IEEE ICOIN Best Paper Award (2021), IEEE Vehicular Technology Society (VTS) Seoul Chapter Awards (2019, 2021, and 2022), and IEEE ICTC Best Paper Award (2022). 
\end{IEEEbiography}
\end{document}